\documentclass[11pt,a4paper,reqno]{amsart}

\usepackage{amsmath,amssymb}
\usepackage{graphicx}
\usepackage[pagebackref, colorlinks = true, linkcolor = blue, urlcolor  = blue, citecolor = red]{hyperref}
\usepackage{xcolor}
\usepackage[margin=1in]{geometry}
\usepackage{caption} 
\usepackage{subcaption} 
\usepackage{graphbox}
\usepackage{array}

\usepackage{listings}
 \definecolor{darkgreen}{rgb}{0.0, 0.2, 0.13}

\newtheorem{theorem}{Theorem}[section]
\newtheorem*{theorem*}{Theorem}
\newtheorem{definition}[theorem]{Definition}

\DeclareMathOperator{\Tr}{Tr}

\DeclareMathOperator{\id}{id}
\DeclareMathOperator{\Wg}{Wg}
\newcommand{\cA}{\mathcal{A}}
\newcommand{\cZ}{\mathcal{Z}}
\newcommand{\cS}{\mathcal{S}}

\begin{document}

\date{\today}

\title[RTNI -- A symbolic integrator for Haar-random tensor networks]{RTNI -- A symbolic integrator for Haar-random\\
tensor networks}

%Random Tensor Network Intergrator: a symbolic package for unitary group integrals}

\author{Motohisa Fukuda}
\address{MF: Yamagata University, 1-4-12 Kojirakawa, Yamagata, 990-8560 Japan}
\email{fukuda@sci.kj.yamagata-u.ac.jp}

\author{Robert K\"onig}
\address{RK: Institute for Advanced Study \& Zentrum Mathematik, Technical University of Munich, Garching, Germany}
\email{robert.koenig@tum.de}

\author{Ion Nechita}
\address{IN: Laboratoire de Physique Th\'eorique, Universit\'e de Toulouse, CNRS, UPS, France
}
\email{nechita@irsamc.ups-tlse.fr}

\subjclass[2000]{}

\begin{abstract}
We provide a  computer algebra package called Random Tensor Network Integrator (\texttt{RTNI}). It allows to compute averages of tensor networks containing multiple Haar-distributed random unitary matrices and deterministic symbolic tensors.  Such tensor networks are represented as multigraphs, with vertices corresponding to tensors or random unitaries and edges corresponding to tensor contractions. Input and output spaces of random unitaries may be subdivided into arbitrary tensor factors, with dimensions treated symbolically. The algorithm implements the graphical Weingarten calculus and produces a weighted sum of tensor networks representing the average over the unitary group. We illustrate the use of this algorithmic tool on some examples  from quantum information theory, including entropy calculations for random tensor network states  as considered in toy models for holographic duality. Mathematica and Python implementations are supplied.  

\end{abstract}

\maketitle

\tableofcontents

\section{Introduction}

The probabilistic method -- made famous by Paul Erd\"os~\cite{erdoes47}  -- now pervades all of discrete mathematics and computer science. In its most well-known form, it is used to demonstrate the existence of certain combinatorial objects by defining a suitable probability space, and showing that randomly chosen elements of that space exhibit the required property with non-zero probability.  A prime example of this strategy is the proof of the achievability part of the channel coding theorem by Shannon~\cite{shannon48} in classical information theory, established roughly at the same time: here the method is applied to show the existence of capacity-achieving codes for a given communication channel. To mention just one additional famous example beyond the original applications to graph theory, the probabilistic method has been used to estimate sizes of $\varepsilon$-nets (respectively $\varepsilon$-samples) in spaces with limited Vapnik-Chervonenkis-Dimension~\cite{vcdim71}.

While the probabilistic method constitutes a powerful mathematical tool, much more direct manifestations of randomness in the real world have been studied throughout the history of physics. Here we are particularly interested in the appearance of random matrices. The consideration of random matrices was initiated by Wigner in order to explain universal features of the eigenvalue spacing of atomic spectra~\cite{wigner55}, but has since then gone undergone significant developments -- see e.g.,~\cite{guhr98} for an in-depth review. Significant fields of application include the theory of disorder and associated transport phenomena in condensed matter systems, and the development of random matrix models in quantum field theory. 

Random matrices, and more specifically the Haar distribution over a compact Lie group, also play a central role in quantum mechanics because of conservation laws. For example, any physical process respecting certain superselection rules can be described by a quantum channel (a completely positive trace-preserving map) which is covariant under the relevant group action. Thus such a channel can be seen as the result of ``supertwirling'', i.e., the consequence of averaging (the inputs and outputs of) a given channel over the unitary group. Because this averaging operation typically leads to a drastic reduction of the number of free parameters, the set of covariant operations (and states) has often been a fruitful testbed and source of examples, e.g., in entanglement theory~\cite{werner1989quantum}. Covariant channels (or instruments) also appear e.g., when considering certain operational figures of merit which are related for example to average fidelities (with averages taken over the Haar measure). In certain cases, the 
associated optimization problem can then be restricted to operations respecting the underlying symmetry. For example, optimal cloning devices where constructed following this approach in~\cite{wernercloning98,keylwernercloning99}. We note that these are only some of the pioneering contributions to this area.

Interestingly, random matrices are fundamental to quantum information theory even beyond such symmetry considerations. Indeed, it was found that the probabilistic method can  be used to establish existence (and it some cases ``genericity'') of states and operations with particular properties; here the distributions are typically defined in terms of Haar-random unitaries.  Pioneering work in this direction~\cite{Hayden2004,Haydenleungwinter2006} has shown for example the existence and ubiquity of states with a large amount of ``locked'' classical correlations. The probabilistic method was also used by Hastings~\cite{hastings09} to
disprove the additivity conjecture for the Holevo-quantity: 
following a proof strategy successfully applied for R\'enyi entropies earlier~\cite{hayden2008counterexamples,Cubitt2008}
he demonstrated the existence of a pair of quantum channels for which the Holevo-quantity (which determines the classical capacity) is non-additive. Closer to the spirit of Shannon's achievability theorem, the probabilistic method has also been used to show the achievability of the quantum capacity (more precisely, the coherent information): here the argument involves picking a random subspace respectively vectors (according to the Haar measure)  of a Hilbert space~\cite{haydenhorodeckiwinteryard08,haydenshorwinter08}. The corresponding ``decoupling approach'' is now a standard  procedure in quantum information theory. 

We refer to~\cite{collinsnechita16} for a detailed review of applications of random matrix techniques in quantum information theory.  In most of these  applications, averages are  taken over the unitary group on some finite-dimensional space. Typically, one is interested in the moments of some expression (such as a matrix product) involving instances of random unitaries, and these are evaluated in an ad-hoc manner. In recent applications, random tensor network states are considered: here Haar-random unitaries are contracted together with fixed tensors in a network. An example is the consideration of random matrix product states~\cite{Collins2013}, and, more recently, the contraction of 
higher-dimensional random tensor networks in the context of holographic duality~\cite{hayden2016holographic}  (see Example~\ref{sec:randomTNSholographic} below). In these cases, the evaluation of Haar averages of quantities of interest  becomes  highly non-trivial, either because higher moments are required, and/or because the average is over a $n$-tuple of independently and identically distributed unitaries.

 \section{Our contribution}
 Our work is motivated by the ubiquity  of Haar-random averages in a large variety of settings, and the fact that the associated computations are -- on a conceptual level -- very similar. Indeed, the natural language capturing these problems is that of random tensor networks: these allow to capture a variety of 
 mathematical objects/expressions involving random matrices, including matrix products, tracial expressions, and tensor contractions. Furthermore, expectation values of such random tensor networks are themselves linear combinations of tensor networks (which may represent scalars, matrices or tensors), and -- for Haar-random unitaries -- these expectation values can be computed algorithmically using the Weingarten calculus.
 
Even though detailed derivations of the Weingarten calculus can be found in the literature, there appears to be no general implementation relying on symbolic manipulation of tensor networks. Our {\bf Random Tensor Network Integrator (\texttt{RTNI})} package\footnote{Code available at \url{https://github.com/MotohisaFukuda/RTNI}} aims to fill this gap by providing routines for computing expectation values of tensor networks involving random unitaries. This is envisioned to facilitate the study and application of objects involving Haar-random unitaries.

Let us briefly mention existing similar software packages dealing with Weingarten integration we are aware of. The \texttt{IntU} package \cite{puchala2017symbolic} is a \texttt{Mathematica} package used for the computation of the averages monomials in the entries of a random unitary matrices, see also \url{https://zksi.iitis.pl/wiki/projects:intu}. There are two main differences with our implementation. Most importantly, we allow for symbolic Hilbert space dimensions, where as in \texttt{IntU} these dimensions must be fixed numerical integers. Moreover, our input data is a graph, as opposed to a monomial for \texttt{IntU}; we consider our input format more practical, especially for applications of the Weingarten calculus (and additionally provide subroutines for monomials). The \texttt{IntHaar} package \cite{ginory2016weingarten} is a \texttt{MAPLE} implementation for the Weingarten calculus, see \url{http://sites.math.rutgers.edu/~ag930/Maple%20Packages/IntHaar.txt} for the code. 
Similarly to the \texttt{IntU} package, the program computes averages of polynomials in the entries of the random unitary matrix, but the Weingarten functions are computed symbolically. Moreover, the case of Haar-distributed orthogonal and symplectic matrices is implemented.
 
%\textcolor{red}{MAPLE PROGRAM: IS this also for fixed dimensions? If so, add a remark}

\subsubsection*{Outline}
This document describes the \texttt{RTNI} package. In Section~\ref{sec:mathematicalbackground}, we briefly review the underlying Weingarten formula. In Section~\ref{sec:graphical-calculus}, we recall the graphical Weingarten calculus which is realized by the \texttt{RTNI} package. Section~\ref{sec:rtni} discusses the main routines provided by \texttt{RTNI} and their use. Finally, in Section~\ref{sec:examples}, several explicit example computations are provided, illustrating the use of the package.

\section{Mathematical background: the Weingarten formula\label{sec:mathematicalbackground}}
Here we briefly review the definition of some relevant mathematical objects, including, in particular, the Weingarten function. We defer the discussion of the graphical Weingarten calculus to Section~\ref{sec:graphical-calculus} and instead begin with an ``element-wise'' expression of the Weingarten formula. This shows how to compute the average of monomials composed of matrix elements of random unitary matrices with respect to the Haar measure. This problem  was first considered in the physics literature by Weingarten~\cite{weingarten1978asymptotic} in the asymptotic limit of large dimension. An explicit expression for fixed matrix dimensions was obtained in~\cite{collins2003moments} and \cite{collins2006integration} by  Collins, respectively Collins and {\'S}niady. The main insight exploited in those works is that the average of tensor product operators of the form $U^{\otimes n} \otimes \bar U^{\otimes n}$ is a weighted sum of permutation operators. This is a consequence of Schur-Weyl duality. 
\begin{theorem}
\label{thm:wg}
 Let $d,p$ be positive integers and
$i=(i_1,\ldots ,i_p)$, $i'=(i'_1,\ldots ,i'_p)$,
$j=(j_1,\ldots ,j_p)$, $j'=(j'_1,\ldots ,j'_p)$
be $p$-tuples of positive integers from $\{1, 2, \ldots, d\}$. Then
\begin{align}
\label{wgf} \int_{\mathcal U(d)}U_{i_1j_1} \cdots U_{i_pj_p}
\bar U_{i'_1j'_1} \cdots
\bar U_{i'_pj'_p} \mathrm{d}U=\!\!\!
\sum_{\alpha, \beta\in \mathcal S_{p}}\delta_{i_1i'_{\alpha (1)}}\ldots
\delta_{i_p i'_{\alpha (p)}}\delta_{j_1j'_{\beta (1)}}\ldots
\delta_{j_p j'_{\beta (p)}} \Wg_d (\alpha^{-1}\beta)\ .
\end{align}
If $p\neq p'$ then
\begin{equation} \label{eq:Wg_diff} \int_{\mathcal U(d)}U_{i_{1}j_{1}} \cdots
U_{i_{p}j_{p}} \bar U_{i'_{1}j'_{1}} \cdots
\bar U_{i'_{p'}j'_{p'}}\mathrm{d}U= 0\ .
\end{equation}
Here we denoted by $\mathcal U(d)$ the unitary group acting on an $d$-dimensional Hilbert space, and the integrals are taken with respect to the normalized Haar measure on~$\mathcal U(d)$. The function $\Wg_d$ is called the {\em unitary Weingarten function}, see Definition~\ref{def:weingartenfct} below.
\end{theorem}
We note that software packages such as the \texttt{IntU} package \cite{puchala2017symbolic} mentioned above primarily rely on expression~\eqref{wgf}. In contrast, our \texttt{RTNI} package  exploits the graphical calculus introduced in the next section, including, in particular, expression~\eqref{eq:eud} below.

The unitary Weingarten function $\Wg_d(\cdot)$ is a combinatorial object defined as follows.
Let~$\cA(\cS_p)$ be the algebra of complex-valued functions on the symmetric group~$\cS_p$ with product given by the convolution
\begin{align}
(f*g)(\pi)&=\sum_{\tau\in \cS_p} f(\tau)g(\tau^{-1}\pi)\qquad\textrm{ for }f,g\in \cA(\cS_p)\textrm{ and }\pi\in \cS_p\ .
\end{align}
The Dirac-Delta function $\delta_e(\pi)=\delta_{e,\pi}$ for $\pi\in\cS_p$ is the identity element in~$\cA(\cS_p)$. 
For any $z\in\mathbb{C}$, let $h_d\in\cA(\cS_p)$ be defined 
\begin{align}
h_d(\pi)&=d^{\#(\pi)}\qquad\textrm{ for }\pi\in\cS_p\ ,\label{eq:hddefinitionx}
\end{align}
where $\#(\pi)$  denotes the number of cycles of the permutation~$\pi$.  Then $h_d$ belongs to the center~$\cZ(\cA(\cS_p))$. 

\begin{definition}
\label{def:weingartenfct}
The unitary Weingarten function 
\begin{align}
\Wg_d:\mathcal{S}_p & \rightarrow \mathbb{R}\\
\sigma & \mapsto \Wg_d(\sigma)
\end{align}
is a function of a dimension parameter $d\in\mathbb{C}$ and of a permutation $\sigma\in\mathcal S_p$. It is defined as
the pseudo-inverse of the element $h_d\in\cA(\cS_p)$ defined in~\eqref{eq:hddefinitionx}, i.e., the unique element in $\cZ(\cA(\cS_p))$ satisfying
\begin{align}
h_d*\Wg_d*h_d&=h_d\qquad\textrm{ and }\qquad \Wg_d*h_d*\Wg_d=\Wg_d\ .
\end{align}
\end{definition}
It can be shown that if $d\not\in \{0,\pm 1,\ldots,\pm (p-1)\}$, then $\Wg_d$ and $h_d$ are inverses of each other
\begin{align}
\sum_{\tau\in\cS_p} \Wg_d(\tau)d^{\#(\tau^{-1}\pi)}&=\delta_{e,\pi}\qquad\textrm{ for all }\pi\in \cS_p\ .
\end{align} 
Furthermore, for integer $d\geq p$, the Weingarten functions can be written as 
\begin{align}\label{eq:wg}
\Wg_d(\sigma) = \frac 1{(p !)^2} \sum_{\lambda \vdash p} \frac{(\chi^{\lambda}(e))^2}{s_{\lambda,d}(1)}\chi^{\lambda}(\sigma)
\end{align}
Here, $\lambda \vdash p$ means that $\lambda$ is a partition of the integer $p$, $\chi^{\lambda}$ is the character of the irreducible representation of the symmetric group $\mathcal S_p$ specified by~$\lambda$,
and $s_{\lambda,d}(1)$ is the Schur polynomial evaluated at the identity (i.e., the
dimension of the irreducible representation of $\mathcal{U}(d)$ with highest weight~$\lambda$. The latter is given by the formula
\begin{align}\label{eq:schur}
s_{\lambda,d} (1) = \prod_{1 \leq i < j \leq d} \frac{\lambda_i - \lambda_j + j -i}{i-j}\ .
\end{align}

For integer $d<p$, the element $h_d\in\cA(\cS_p)$ has no inverse, and formula~\eqref{eq:wg}  no longer applies, but a separate formula is needed. Indeed, expression~\eqref{eq:wg} has poles when~$d<p$. 
However, in computing averages over the Haar measure as in Theorem~\ref{thm:wg}, those
 poles cancel, yielding a rational function in $d$ on the RHS~of~\eqref{wgf}. This is why one can use the expression of Eq.~\eqref{eq:wg} for the Weingarten function~$\Wg_d$ for all $d\geq 1$ in the computations as in Theorem~\ref{thm:wg}. We refer to~\cite{collins2006integration} for more details.

\section{Graphical notation for tensors and for the Weingarten formula}
\label{sec:graphical-calculus}
In this section, we briefly recall the graphical calculus method for computing unitary integrals introduced in \cite{collins2010random}, to which the readers are referred for more details. As discussed below,  the \texttt{RTNI} package provides a general  implementation of this method.

First, we explain the graphical formalism for tensor calculus, which can be traced back to Penrose \cite{penrose1971applications}; for modern presentations, see \cite{collins2010random} or \cite{coecke2010quantum}.
The main object here is a complex vector space of diagrams. An element of this vector space will be represented as a list of pairs of the form $(g_i,w_i)$, where $g_i$ is a graph (a.k.a.~a diagram or tensor network) and $w_i$ is a scalar weight; the corresponding element is $\mathcal D = \sum_i w_i  \cdot g_i$. The diagrams (or graphs) correspond to tensors, and consist of \emph{boxes} and \emph{wires}. 
Boxes in a graph represent tensors. Each box has attached to it symbols of different shapes, where each shape corresponds to a vector space; identical shapes depict isomorphic vector spaces.  
Those symbols are  filled (black) or empty (white), corresponding to primal or dual spaces. 
Wires connect these symbols, corresponding to tensor contractions $V \times V^* \to \mathbb C$. This is why each wire must connect symbols of the same shape, i.e.~the same dimension. 
A diagram is a collection of such boxes and wires and corresponds to an element in a tensor product space.  If there is more than one connected component in a graph, then the components are multiplied by taking the tensor product.

\begin{figure}[htbp]
\includegraphics[scale=1]{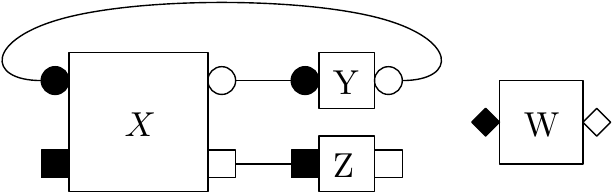}
\caption{A diagram  for $\Tr_1[ X (Y \otimes Z)] \otimes W$. The matrices $X,Y,Z,W$ act on the spaces $\mathbb C^k \otimes \mathbb C^n, \mathbb C^k, \mathbb C^n, \mathbb C^d$ respectively. Round decorations correspond to~$\mathbb C^k$, square decorations to~$\mathbb C^n$, and diamond-shaped decorations correspond to~$\mathbb C^d$.}
\label{fig:ed}
\end{figure}

For example, the tensor (here, a bipartite matrix) $\Tr_1[ X (Y \otimes Z)] \otimes W$, where $\Tr_1$ stands for the partial trace over the first space, is represented in Figure~\ref{fig:ed}. Consider the primal side of the box for the  matrix $X$. There are two symbols of different shapes implying that the  effective primal space of $X$ is the tensor product of these two spaces. 

Next, we describe how to efficiently compute expected values of diagrams which contain boxes of random unitary matrices. This technique in fact can be extracted from Theorem \ref{thm:wg}, once one uses the fact that the averaging process is linear.
More precisely, the delta functions in each term in the RHS of~\eqref{wgf} simply indicate how one needs to reconnect boxes corresponding to the random unitary operator $U$ and $\bar U$. For each pair of permutations $(\alpha,\beta)$ in  \eqref{wgf}, one eliminates $U$ and $\bar U$ boxes and reconnects the wires originally connected to these boxes to get a new diagram. Since these manipulations are made on wires and symbols (boxes), one can regard them as operations on a graph, identifying wires and symbols with edges and vertices, respectively. 
This process for a fixed pair of permutations is called a removal and the whole process which sums all the new graphs  over all permutations is called the graph expansion. Formally, if $\mathcal D$ is a diagram containing boxes corresponding to a Haar-distributed random unitary matrix $U$, then its expectation value (with respect to~$U$) is 
\begin{align}
\mathbb{E}_U(\mathcal{D})=\sum_{\alpha,\beta} \mathcal{D}_{\alpha,\beta}\Wg_d(\alpha^{-1}\beta)\ ,\label{eq:eud}
\end{align}
where $\mathcal D_{\alpha,\beta}$ is the diagram obtained from $\mathcal D$ by removing the boxes associated with $U$ and $\bar U$ and adding extra wires to the resulting diagram, as follows: we connect white decorations of the $i$-th $U$ box with the white decorations of the $\alpha(i)$-th $\bar{U}$ box and, by a similar procedure, the black decorations are paired using the $\beta$ permutation. One can see how this works in examples in Section \ref{sec:examples}, see in particular Figures \ref{fig:XUYUstar} and \ref{fig:partial-trace}.

\section{The \texttt{RTNI} package\label{sec:rtni}}
In this section, we explain how to use the \texttt{RTNI} package, available at \url{https://github.com/MotohisaFukuda/RTNI}. Readers are also referred to some examples in Section~ \ref{sec:examples} while reading this section.

The following two functions are the main routines provided by \texttt{RTNI}:
\begin{enumerate}
	\item $\textbf{integrateHaarUnitary}(TNList,\ varName,\ inDims,\ outDimns,\ totalDims)$: integrates out the random unitary matrix $varName$ in the tensor networks in $TNList$. 
	\item $\textbf{visualizeTN}(TNList,\{EdgeLabeling\rightarrow True\})$: produces a graphical output representing each tensor network (graph) in the list, together with the associated weights.
\end{enumerate}
In Sections~\ref{sec:int} and~\ref{sec:vis}, we provide a detailed explanation of the functionality 
realized by these routines, as well as the syntax used for inputs and outputs.  These two routines 
are the only routines needed for working with random tensor networks.

We also provide an additional auxiliary routine for matrix products, which covers many simple use-cases.  It internally creates a tensor network representing the matrix product (respectively its trace), invokes \textbf{integrateHaarUnitary}, and 
re-expresses the resulting linear combination of tensor networks in terms of matrix products/tracial expressions.
\begin{enumerate}\setcounter{enumi}{2}
	\item $\textbf{MultinomialexpectationvalueHaar}(Dim,\ eList,\ variableList,\ usetrace)$: symbolically computes moments of matrix products involving Haar-random unitaries. 
\end{enumerate}
The corresponding syntax is explained in Section~\ref{sec:multinomialexp}.

\subsection{integrateHaarUnitary}\label{sec:int}
An input of \textbf{integrateHaarUnitary} specifies the diagrams (tensor networks) containing random matrices to be averaged. At the program level, inputs of the function \textbf{integrateHaarUnitary} divide into five variables as indicated above. In the following, we describe each of them.
\begin{enumerate}
\item {\bf TNList} is a  list of pairs~(tns$_j$,weight$_j$)  represeting a weighted linear combination of~$n$ (with $n\in\mathbb{N}$ arbitrary) tensor networks,
where the $j$-th tensor network~tns$_j$ has weight weight$_j$. The syntax is as follows:
\begin{quote}
\vspace{1ex}
\{ \{ tns$_1$, weight$_1$ \}, \{ tns$_2$, weight$_2$ \},\ldots, \{ tns$_n$, weight$_n$ \} \}
\vspace{1ex}
\end{quote}
%\vspace{1ex}
%\begin{itemize} 
%\item \Mathematica: \{ \{ tns$_1$, weight$_1$ \}, \{ tns$_2$, weight$_2$\},\ldots, \{tns$_n$, weight$_n$\} \}
%\item \Python:   [ [ tns$_1$, weight$_1$ ], [ tns$_2$, weight$_2$],\ldots, [tns$_n$, weight$_n$] ]
%\end{itemize}
%\vspace{1ex}
For $n=1$ (i.e., if only one tensor network is considered) and the associated weight is equal to~$1$, the alternative expression
%\vspace{1ex}
%\begin{itemize} 
%\item \Mathematica:  tns
%\item \Python:   tns
%\end{itemize}
%\vspace{1ex}
\begin{quote}
\vspace{1ex}
tns
\vspace{1ex}
\end{quote}
may be used. In these expressions, each term ``tns'' is a list of edges of the form
\begin{quote}
\vspace{1ex}
\{ \{ vertex, vertex \}, \ldots, \{ vertex, vertex \} \} 
\vspace{1ex}
\end{quote}
%\vspace{1ex}
%\begin{itemize}
%\item
%\Mathematica: \{ \{ vertex, vertex \}, \ldots, \{ vertex, vertex \} \} 
%\item \Python:  [ [ vertex, vertex ], \ldots, [ vertex, vertex ] ]
%\end{itemize}
\vspace{1ex}
and ``weight'' is a symbolic expression or scalar specifying the weight of the associated tensor network.

Vertices correspond to the different shaped and colored decorations attached to boxes in the graphical calculus from Section \ref{sec:graphical-calculus}. Importantly, each term ``vertex''  is given by a list of four items  in the format
\begin{quote}
\vspace{1ex}
\{varName, ID, IN/OUT, leg\}
\vspace{1ex}
\end{quote}
%\vspace{1ex}
%\begin{itemize}
%\item
%\Mathematica: \{varName, ID, IN/OUT, leg\}
%\item \Python:  [varName, ID, IN/OUT, leg]\ 
%\end{itemize}
%\vspace{1ex}
 where
\begin{itemize}
	\item varName is a string denoting the name of the corresponding box; for boxes associated to Haar random unitary matrices, this string can end in the star symbol ``$*$'', in which case this references the adjoint of the unitary operator.
	\item ID is an integer used to distinguish between several copies of the same box; this is in particular used to distinguish different copies of random unitary matrices and their adjoints.
	\item IN/OUT is a  string, referring to whether the vertex corresponds to an input (''in'') or an output (''out'') of a matrix. 
	\item leg is a positive integer corresponding to which tensor factor the vertex corresponds to.
\end{itemize}
We display in Figure \ref{fig:edges-vertices} an example  of a tensor network; the three edges appearing are encoded by (from left to right, and top to bottom; the colors are used for clarity only):
\begin{align*}
\textcolor{darkgreen}{e_1} &= \{\{"U", \textcolor{blue}{1}, "out", 1\},\{"U", \textcolor{blue}{2}, "out", 1\}\}\\
\textcolor{darkgreen}{e_2} &= \{\{"U", \textcolor{blue}{1}, "in", 1\},\{"A", \textcolor{blue}{1}, "out", \textcolor{red}{1}\}\}\\
\textcolor{darkgreen}{e_3}&= \{\{"U", \textcolor{blue}{2}, "in", 1\},\{"A", \textcolor{blue}{1}, "out", \textcolor{red}{2}\}\}.
\end{align*}
\begin{figure}
	\centering
	\includegraphics[width=0.4\linewidth]{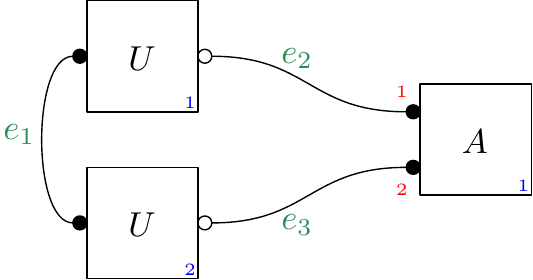}
	\caption{A tensor network  containing three edges (wires). The $U$-boxes on the left have IDs \textcolor{blue}{1} and \textcolor{blue}{2} (labels in the bottom-right corner of the boxes), while $A \in V^{(\textcolor{red}{1})} \otimes V^{(\textcolor{red}{2})}$
 occurs only once (and the ID~\textcolor{blue}{1} is used) and has two legs numbered~\textcolor{red}{1} and~\textcolor{red}{2}, respectively (labels near the decorations of the $A$ box).   Output vertices are marked by filled circles, whereas input vertices are marked by empty circles. The edges also carry labels, to make their encoding more clear.}
	\label{fig:edges-vertices}
\end{figure}

\item {\bf varName} is a letter showing over which random matrix the tensor network (respectively linear combination of tensor networks) is to be averaged.
\item {\bf inDims} is a list specifying the dimensions of the factors in the tensor product corresponding to the outputs of the random unitary matrix.
\item {\bf outDims} is a list specifying the dimensions of the factors in the tensor product corresponding to the inputs of the random unitary matrix.  
\item {\bf totalDims} is the size of the random unitary matrix to be integrated out.
\end{enumerate}
It is important to note that the elements of the lists {\bf outDims}, {\bf inDimns}, as well as {\bf totalDims} can be symbolic. 

An output of \textbf{integrateHaarUnitary} is a list made of pairs of tensor networks and weights:
\begin{quote}
\vspace{1ex}
\{ \{ tns$_1$, weight$_1$ \}, \{ tns$_2$, weight$_2$ \},\ldots, \{ tns$_m$, weight$_m$ \} \}
\vspace{1ex}
\end{quote}
%\vspace{1ex}
%\begin{itemize} 
%\item \Mathematica: \{ \{ tns$_1$, weight$_1$ \}, \{ tns$_2$, weight$_2$\},\ldots, \{tns$_n$, weight$_n$\} \}
%\item \Python:   [ [ tns$_1$, weight$_1$ ], [ tns$_2$, weight$_2$],\ldots, [tns$_n$, weight$_n$] ]
%\item \Mathematica: \{ \{ \{ edgeList, \ldots, edgeList \}, weight \}, \ldots, \{ \{ edgeList, \ldots, edgeList \}, weight \} \}
%\item \Python: [ [ [ edgeList, \ldots, edgeList ], weight  ], \ldots, [ [ edgeList, \ldots, edgeList ], weight  ] ]
%\end{itemize}
%\vspace{1ex}
Hence the output can be fed into  \textbf{integrateHaarUnitary} for iterative calculations.
Note that  \textbf{integrateHaarUnitary} creates new edges (and new vertices): these correspond to symbols on the random matrix which are not connected to other symbols (which did not appear in \textbf{TNList}). In the \texttt{Mathematica} package the names of such new vertices starts with ``dummy'', and in the  \texttt{Python} package with ``@''. One can also add such extra vertices  manually.

\subsection{visualizeTN}\label{sec:vis}
\begin{figure}
	\centering
	\includegraphics[width=0.4\linewidth]{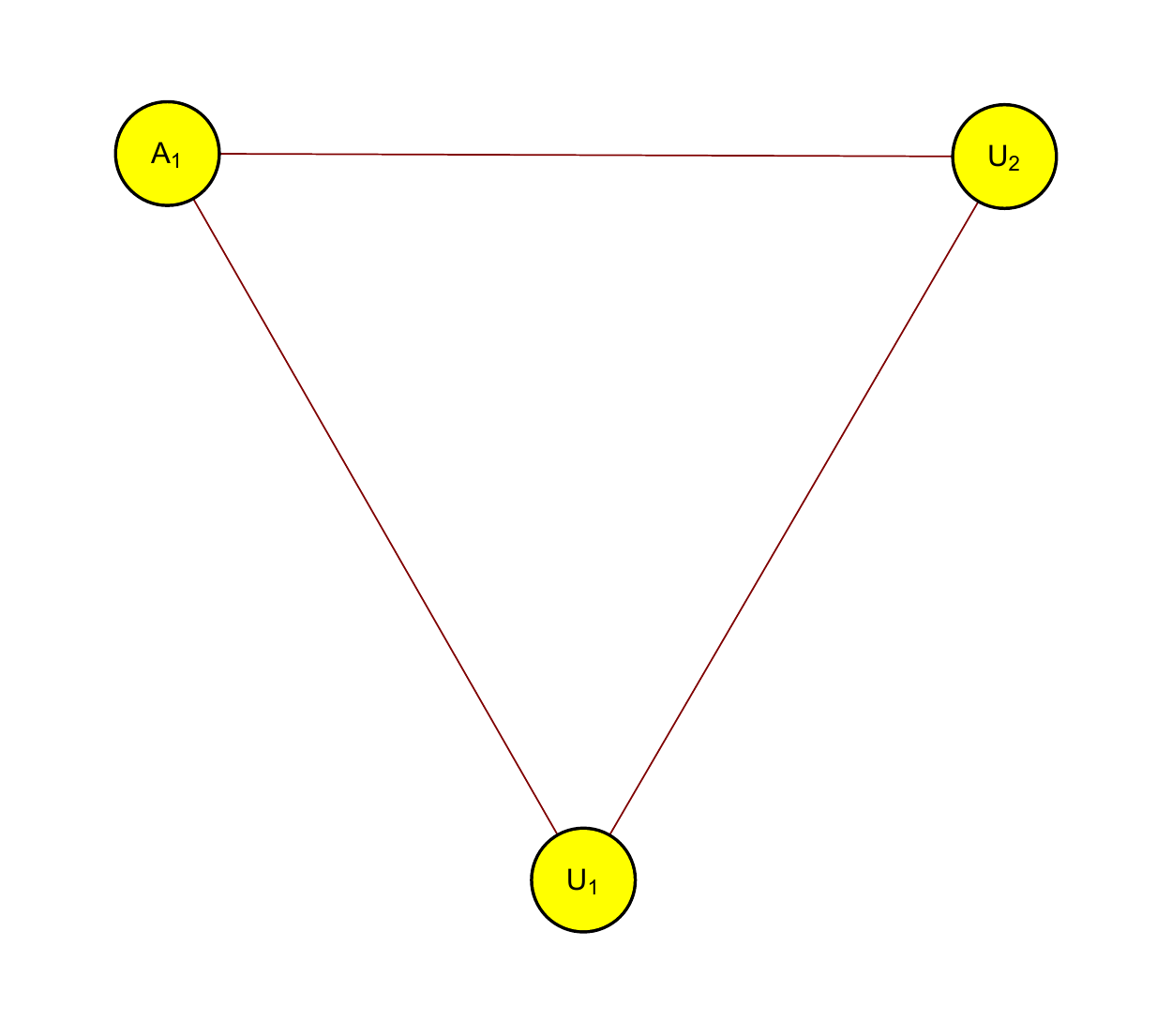}	\includegraphics[width=0.4\linewidth]{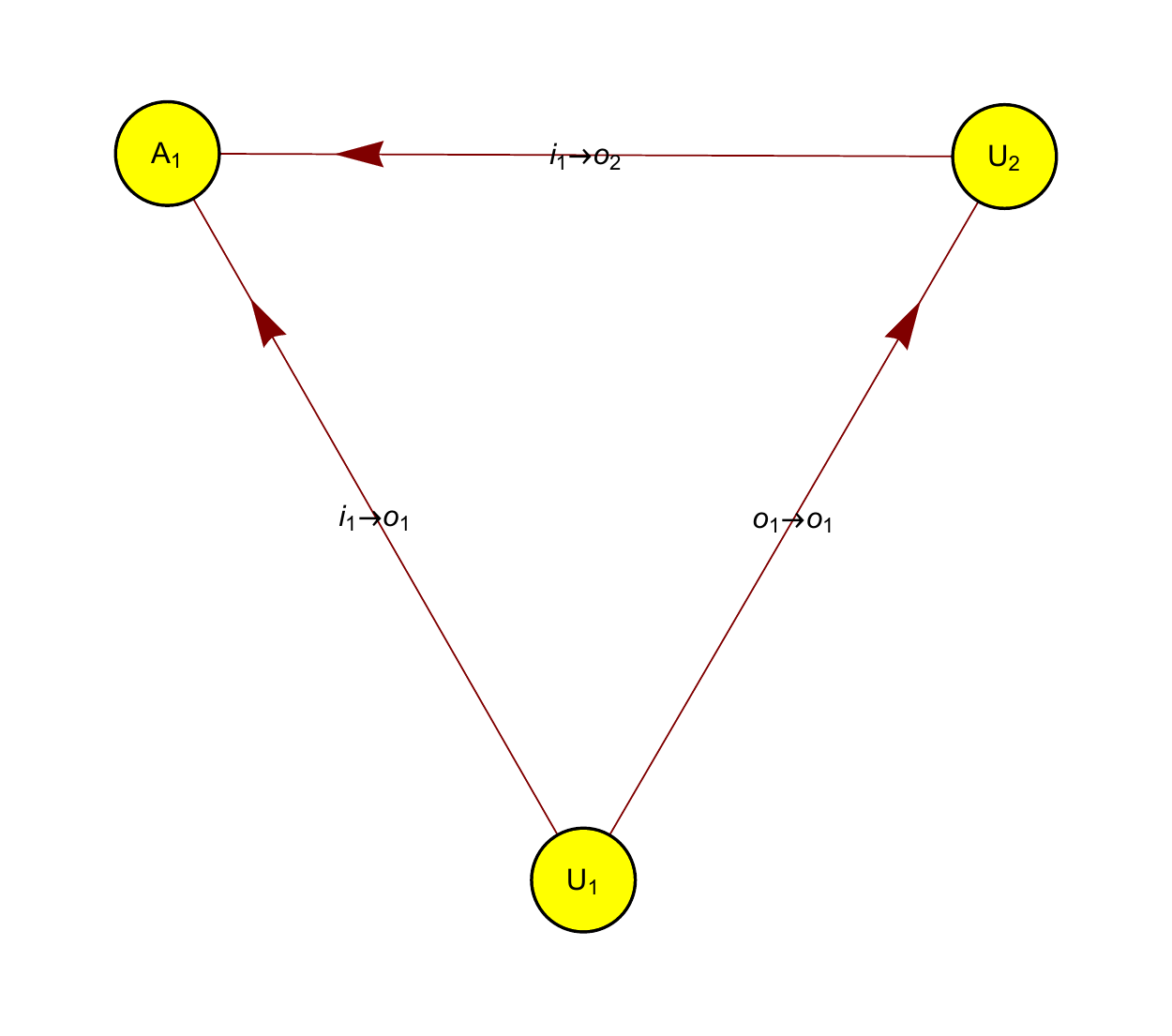}
	\caption{The tensor network from Fig.~\ref{fig:edges-vertices}, as visualized by
	the routine \textbf{visualizeTN}, with edge labels turned off (left) respectively shown (right)
	The commands used to produce these figures are $\textsf{visualizeTN[\{e1,e2,e3\}]}$ and $\textsf{visualizeTN[\{e1,e2,e3\},\{EdgeLabeling$\rightarrow $True\}]}$, respectively.\label{fig:visualizeTN}
	}
\end{figure}

As the name  indicates, this function creates a graphical representation of a single tensor network respectively a weighted list of tensor networks.  An input of \textbf{visualizeTN} is the first argument of an  input, or an output of \textbf{integrateHaarUnitary}. An example is shown in Fig.~\ref{fig:visualizeTN}. The arguments are as follows:
\begin{enumerate}
\item {\bf TNList} This is either a single tensor network, or a list of (weighted) tensor networks as in the description above. 
\item {\bf \{EdgeLabeling$\rightarrow$ True\}} This optional second argument forces the visualization to display edge labels, indicating which input/output is connected to which input/output. This argument can be omitted, in which case edges are not labeled.
\end{enumerate}

\subsection{MultinomialexpectationvalueHaar}\label{sec:multinomialexp}
An input of \textbf{MultinomialexpectationvalueHaar} 
specifies a ``matrix product'' expression of the form
\begin{align}
X_1U_1X_2U_3\cdots X_nU_n\qquad\textrm{  respectively its trace }\qquad \Tr(X_1U_1X_2U_3\cdots X_nU_n)\ ,\label{eq:ujexpression}
\end{align}
 where each $U_j\in \{U,U^*,U^T,\overline{U}\}$
is either a fixed unitary~$U$, its Hermitian adjoint~$U^*$, its transpose~$U^T$, or its entriwise complex conjugate~$\overline{U}$. The function takes an additional dimension parameter~$d$ and computes the Haar averaged expression symbolically. In more detail, the input variables are the following:
\begin{enumerate}
\item
{\bf Dim} is (a symbolical expression giving) the dimension  of the matrices that are integrated over.
\item
{\bf  eList} is a list of the form $\{\epsilon_1,\ldots,\epsilon_n\}$ (for any integer~$n$). Here $\epsilon_j\in \{1,2,3,4\}$
for each~$j$. Here the values  define the unitary part of the expressions~\eqref{eq:ujexpression}
by the following translation table:
\begin{center}
\begin{tabular}{l|c|c|c|c|}
$\epsilon_j$ & 1 & 2 & 3 & 4\\
\hline
$U_j$         &  $U$ & $U^*$ & $U^T$ & $\overline{U}$
\end{tabular}
\end{center}

\item
{\bf variableList}
is a list of the form $\{X_1,\ldots,X_n\}$. Here the $j$-term directly specifies the $j$-th variable~$X_j$ in an expression of the form~\eqref{eq:ujexpression}
\item
{\bf usetrace} is either $\mathsf{True}$ or $\mathsf{False}$. It determines whether the first or second term in~\eqref{eq:ujexpression} is computed according to the table
\begin{center}
\begin{tabular}{l|c|c|}
usetrace & $\mathsf{True}$ & $\mathsf{False}$\\
\hline
expression computed & $\Tr(X_1U_1X_2U_3\cdots X_nU_n)$ & $X_1U_1X_2U_3\cdots X_nU_n$
\end{tabular}
\end{center}
\end{enumerate}
The output of ``MultinomialexpectationvalueHaar''  is a symbolic expression giving corresponding Haar average. Examples  for this function are given in Section~\ref{sec:multinomialexpectationvalueexamples}.

\subsection{Weingarten functions and precomputed values }
A table of precomputed Weingarten function is loaded upon program initialization. 
These are produced by a program in the \texttt{Python} package, and are stored in a subdirectory called \textsf{precomputedWG}. The latter computes all values of Weingarten functions based on expression~\eqref{eq:wg}. At present, all Weingarten functions with $p\leq 20$ are provided, with $d$~being a symbolic parameter. In particular, this means that the \texttt{Mathematica} package can only compute averages of tensor networks containing up to $p=20$ boxes corresponding to a random unitary matrix of some type (and the corresponding number of conjugate/adjoint boxes).

\subsection{Differences between the \texttt{Mathematica} and \texttt{Python} packages}\label{sec:mp}
There are two implementations for \texttt{Mathematica} and \texttt{Python}, but they work almost identically from the user's viewpoint, with the replacement of curly brackets (\texttt{Mathematica}) with square brackets (\texttt{Python}).

In the \texttt{Mathematica} package, it is possible to give as an input a single diagram, in which case the function assigns it the weight~1, as specified above. This feature is not implemented in \texttt{Python}.

In the \texttt{Python} package, the input of the routine \textbf{integrateHaarUnitary} is of the form (TNList, randomMatrixList), 
where
\begin{enumerate}
\item
\textbf{TNList} is a (weighted) list of tensor networks as above, and 
\item
\textbf{randomMatrixList} is a list of the form
\begin{quote}
\vspace{1ex}
 [ [varName, outDims, inDimns, totalDims], \ldots, [varName, outDims, inDimns, totalDims] ]
 \vspace{1ex}
 \end{quote}
 \end{enumerate}
 The routine then computes the average of a tensor network over multiple random unitary matrices at once. In contrast, in \texttt{Mathematica}, the routine needs to be invoked several times. 
 
 The \texttt{Mathematica} package loads the (provided) precomputed values of the Weingarten functions. The \texttt{Python} routine
first  checks if there is a preexisting set of precomputed Weingarten function for~$p$ as required. If not, this is computed and  saved to the folder named ``Weingarten''. To this end, the Python package also creates folders named ``SGC'' and ``SP'' which contain the characters of symmetric groups and Schur polynomials. 
See Appendix \ref{sec:GW} for more details. 

%\vspace{-0.02ex} % keeps things on the same page

\section{Examples}\label{sec:examples}
Here we present a collection of simple examples of use cases for the package \texttt{RTNI}. We gradually work our way up from basic situations to more involved cases, presenting the code (of the \texttt{Mathematica} implementation) and comparing with analytical computations. 
\subsection{Twirling a matrix}
\begin{figure}[htbp]
	\begin{center}
		\includegraphics[scale=1]{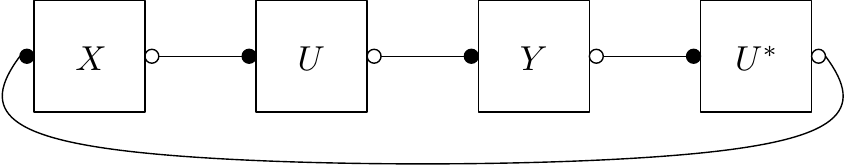} \\
		\vspace{1cm}
		\includegraphics[scale=1]{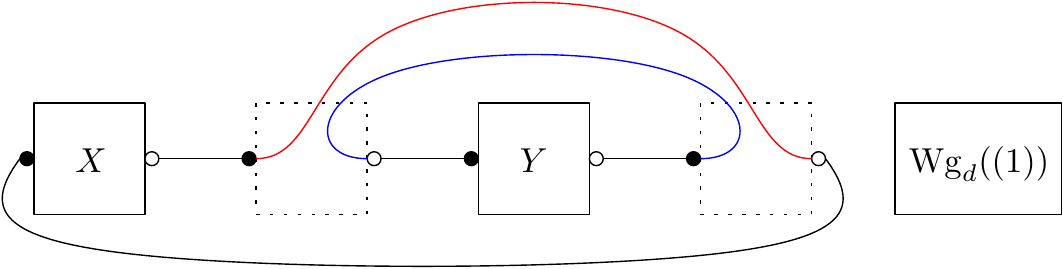}
		\caption{Top: the diagram for $\Tr[XUYU^*]$. Bottom: the average of the diagram on top, over $U \in \mathcal U_d$. The box corresponding to the (scalar) value of the Weingarten function is equal to $1/d$. One recognizes the product of the traces of the matrices $X$ and $Y$.}
		\label{fig:XUYUstar}
	\end{center}
\end{figure}

\begin{figure}[htbp]
	\begin{center}
\includegraphics[align=c, scale=0.6]{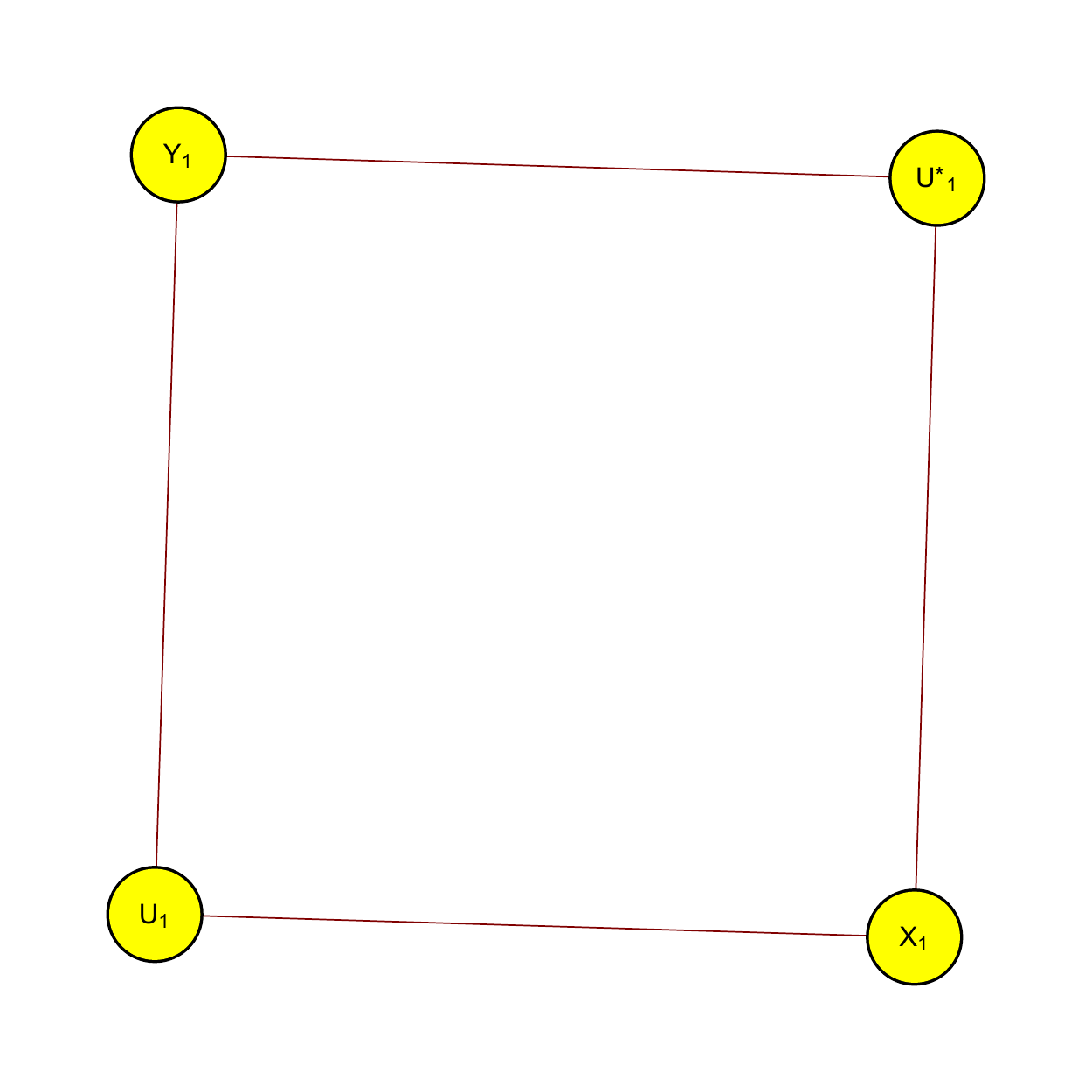} \qquad\qquad\qquad \includegraphics[align=c, scale=0.8]{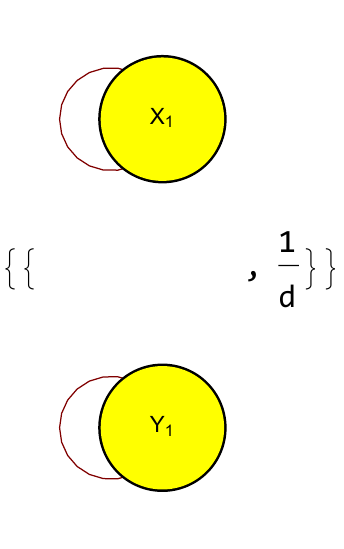}
			\caption{Left: Visualization in \texttt{Mathematica} for the tensor network associated to $\Tr[XUYU^*]$, generated with \textbf{visualizeTN}[g]. Right: The output tensor network after integrating, with respect to the Haar random unitary matrix $U$, the tensor network on the left (figure generated with \textbf{visualizeTN}[Eg]). The result consists of two tensor networks, corresponding respectively to $\Tr X$ and $\Tr Y$, together with the Weingarten weight $1/d$. This
			therefore represents the result~\eqref{eq:vexplicitlycomputed}. 			}
		\label{fig:mTrXUYUstar}
	\end{center}
\end{figure}

We start with the simplest possible example, which is to compute the overlap between a fixed matrix $X$ and a ``twirled'' matrix $UYU^*$, where $Y$ is a fixed $d \times d$ matrix and $U$ is a Haar-distributed random unitary matrix (also of size $d$). The result is a scalar
$$v = \mathbb E \Tr[XUYU^*] = \int_{U \in \mathcal U(d)} XUYU^* \, \mathrm{d}U\ .$$
The diagram corresponding to the integral above, and its Weingarten expansion are depicted in Figure \ref{fig:XUYUstar}; the result is 
\begin{align}
v = \frac{(\Tr X)(\Tr Y)}{d}\ .\label{eq:vexplicitlycomputed}
\end{align}
In order to implement this in the \texttt{RTNI} package, we first need to input the corresponding tensor network (see Figure \ref{fig:XUYUstar} top): it is a square, having 4 edges. 
\begin{lstlisting}
In[1]:= e1 = {{"U", 1, "out", 1}, {"X", 1, "in", 1}};
In[2]:= e2 = {{"Y", 1, "out", 1}, {"U", 1, "in", 1}};
In[3]:= e3 = {{"U*", 1, "out", 1}, {"Y", 1, "in", 1}};
In[4]:= e4 = {{"X", 1, "out", 1}, {"U*", 1, "in", 1}};
In[5]:= g = {e1, e2, e3, e4}
Out[1]= {{{U, 1, out, 1}, {X, 1, in, 1}}, {{Y, 1, out, 1}, {U, 1, in, 1}}, 
        {{U*, 1, out, 1}, {Y, 1, in, 1}}, {{X, 1, out, 1}, {U*, 1, in, 1}}}
\end{lstlisting}
The tensor network described by this code is visualized on the left in Figure \ref{fig:mTrXUYUstar}, using the subroutine~\textbf{visualizeTN}; it is essentially identical to Figure~\ref{fig:XUYUstar}.   The four edges correspond to the four matrix products in the trace. Let us analyze in detail the first edge, $e1$. Its source is the $X$-box having ID 1 (there is only one such box, i.e.~$X$ occurs only once in the diagram), and more precisely it is an ``input'' of that box (hence the ''in'' in the third position of the list describing the source). Since there is just one vector space in this example, the fourth parameter is 1. The sink of $e1$ is the output (the ''out'' in the third position) of the first (the 1 in the second position) $U$ box; again, the last parameter is 1, indicating that we are using the first (and only) factor of the tensor product here. The tensor network $g$ is the collection of the four edges, and it is displayed using the function \textbf{visualizeTN}. Next, we perform the unitary integration, as follows:
\begin{lstlisting}
In[1]:= Eg = integrateHaarUnitary[g, "U", {d}, {d}, d]
Out[1]= {{{{{Y, 1, out, 1}, {Y, 1, in, 1}}, {{X, 1, in, 1}, 
        {X, 1, out, 1}}}, 1/d}}
\end{lstlisting}
The resulting weighted sum of tensor networks (in this case having only one summand) is  then displayed using the routine \textbf{visualizeTN}, see the right subfigure of Figure~\ref{fig:mTrXUYUstar}. There are  two single vertex loop graphs corresponding to the two traces and the Weingarten weight. 

The command \textbf{integrateHaarUnitary} invoked above is the main routine of our package, and it performs the integral of the tensor network $g$ over the unitaries corresponding to the letter $U$ (these are the boxes $U$ and $U^*$), assigning, in order, dimensions to the input spaces of $U$, the output spaces of $U$, and the total dimension of $U$ (here, there is just one input/output space, and all the dimensions are equal to $d$). The output $Eg$ is a (weighted) list of tensor networks (just one in this case), and we recognize in Figure~\ref{fig:mTrXUYUstar} the tensor network with one vertex ($X$) with one loop attached (corresponding to $\Tr X$), the trace of $Y$, and the weight of the tensor network, $1/d$.

\begin{figure}[htbp]
	\begin{center}
\includegraphics[scale=1.0]{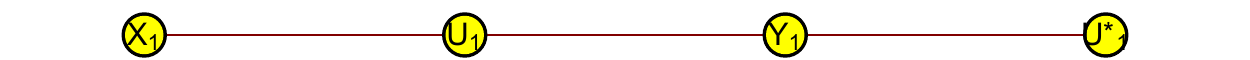} \\ 
\vspace{1cm} \includegraphics[scale=1.0]{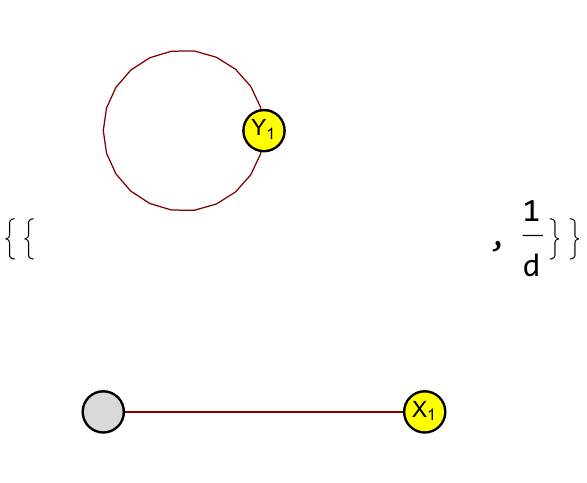}
			\caption{Top: Visualization in \texttt{Mathematica} for the tensor network associated to $XUYU^*$. Bottom: The output tensor network after integrating, with respect to the Haar random unitary matrix $U$, the tensor network on top. The result consists of two tensor networks, corresponding respectively to $X$ and $\Tr Y$, together with the Weingarten weight $1/d$. Observe that there is now dummy vertex attached to~$X$. Thus we recover the result~\eqref{eq:XYddefinition}.}
		\label{fig:mXUYUstar}
	\end{center}
\end{figure}

Next, we consider the same example but without the trace, i.e., the expression~$w = XUYU^*$. Now, the result of the computation is no longer a scalar, but  equal to
\begin{align}
w = \frac{\Tr Y}{d} X\ .\label{eq:XYddefinition}
\end{align} To compute this, the following code can be used:
\begin{lstlisting}
In[1]:= e1 = {{"U", 1, "out", 1}, {"X", 1, "in", 1}};
In[2]:= e2 = {{"Y", 1, "out", 1}, {"U", 1, "in", 1}};
In[3]:= e3 = {{"U*", 1, "out", 1}, {"Y", 1,"in", 1}};
In[4]:= g = {e1, e2, e3}
Out[1]= {{{U, 1, out, 1}, {X, 1, in, 1}}, {{Y, 1, out, 1}, 
        {U, 1,in, 1}}, {{U*, 1, out, 1}, {Y, 1, in, 1}}}
In[5]:= Eg = integrateHaarUnitary[g, "U", {d}, {d}, d]
Out[2]= {{{{{Y, 1, out, 1}, {Y, 1, in, 1}}, 
        {{X, 1, in, 1}, {dummy-U*-IN-1-1, 1, out, 1}}}, 1/d}}
\end{lstlisting}
In this case, the \texttt{RTNI}  package creates  a ``dummy'' vertex for the input of $U^*$, in order to work with a proper tensor network. After the computation, this dummy vertex is connected to the output of $X$, see Figure~\ref{fig:mXUYUstar}.

We note that both expectation values  $\mathbb E \Tr[XUYU^*]$  and $\mathbb{E}XUYU^*$ considered here could alternatively be computed using the subroutine~\textbf{MultinomialexpectationvalueHaar}, see Example~\ref{sec:multinomialexpectationvalueexamples} below. 

\subsection{Several tensor factors}\label{sec:several_tensor}
We now consider an example where the random unitary operators are \emph{bipartite}: they act on a tensor product space. More precisely, we are interested in the partial trace of a twirled operator
$$[\id \otimes \Tr](UAU^*),$$
where the identity operator acts on a tensor factor (with label 1) of dimension $n$, while the trace operator acts on a different tensor factor (with label 2) of dimension $k$.

\begin{figure}[htbp]
	\begin{center}
		\includegraphics[scale=1]{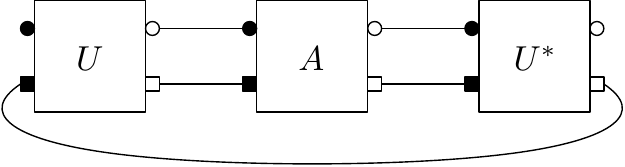} \\
		\vspace{0.5cm}
		\includegraphics[scale=1]{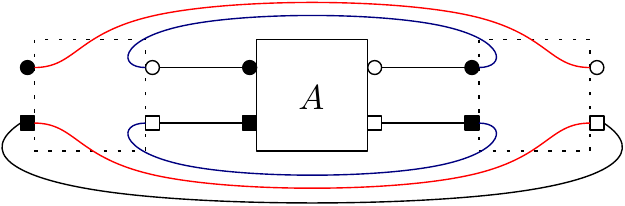}
		\caption{Top: the diagram for $[\id \otimes \Tr](UAU^*)$. Bottom: the average~$\mathbb{E} [\id \otimes \Tr](UAU^*)$ of the diagram on top, over $U \in \mathcal U(d)$, where $d=nk$. Note that the Weingarten weight is omitted here.}
		\label{fig:partial-trace}
	\end{center}
\end{figure}

The analytical computation of the Weingarten integral with respect to $U$ is straightforward: the result reads
\begin{align}
\mathbb E[\id \otimes \Tr](UAU^*) = I_n \Tr(A) k \Wg_{nk}((1)) = \frac{\Tr A}{n} I_n\ .\label{eq:expectationuaustarx}
\end{align} The corresponding tensor networks for both the expresssion~$[\id \otimes \Tr](UAU^*)$ and its expectation are illustrated in Fig.~\ref{fig:partial-trace}.

The \texttt{Mathematica} code for the corresponding tensor network and for the computation of the expectation value with respect to the bipartite unitary matrix $U$ is as follows.  In this example, one needs to specify that the unitary matrices $U$ and $U^*$ act on a tensor product of two vector spaces with respective dimensions $n$ and $k$. The input and output spaces are identical, and the total dimension is $nk$.
\begin{lstlisting}
In[1]:= e1 = {{"A", 1, "out", 1}, {"U", 1,  "in", 1}};
In[2]:= e2 = {{"A", 1, "out", 2}, {"U", 1, "in", 2}};
In[3]:= e3 = {{"U*", 1, "out", 1}, {"A", 1, "in", 1}};
In[4]:= e4 = {{"U*", 1, "out", 2}, {"A", 1, "in", 2}};
In[5]:= e5 = {{"U", 1, "out", 2}, {"U*", 1, "in", 2}};
In[6]:= g = {e1, e2, e3, e4, e5}
Out[1]= {{{A, 1, out, 1}, {U, 1, in, 1}}, {{A, 1, out, 2}, 
        {U, 1, in, 2}}, {{U*, 1, out, 1}, {A, 1, in, 1}}, 
        {{U*, 1, out, 2}, {A, 1, in, 2}}, {{U, 1, out, 2}, 
        {U*, 1, in, 2}}}
In[7]:= Eg = integrateHaarUnitary[g, "U", {n, k}, {n, k}, n k]
Out[2]= {{{{{A, 1, out, 1}, {A, 1, in, 1}}, {{A, 1, out, 2}, 
        {A, 1, in, 2}}, {{dummy-U-OUT-1-1, 1, in, 1},
        {dummy-U*-IN-1-1, 1, out, 1}}}, 1/n}}
\end{lstlisting}

\begin{figure}[htbp]
	\begin{center}
\includegraphics[align=c, scale=0.5]{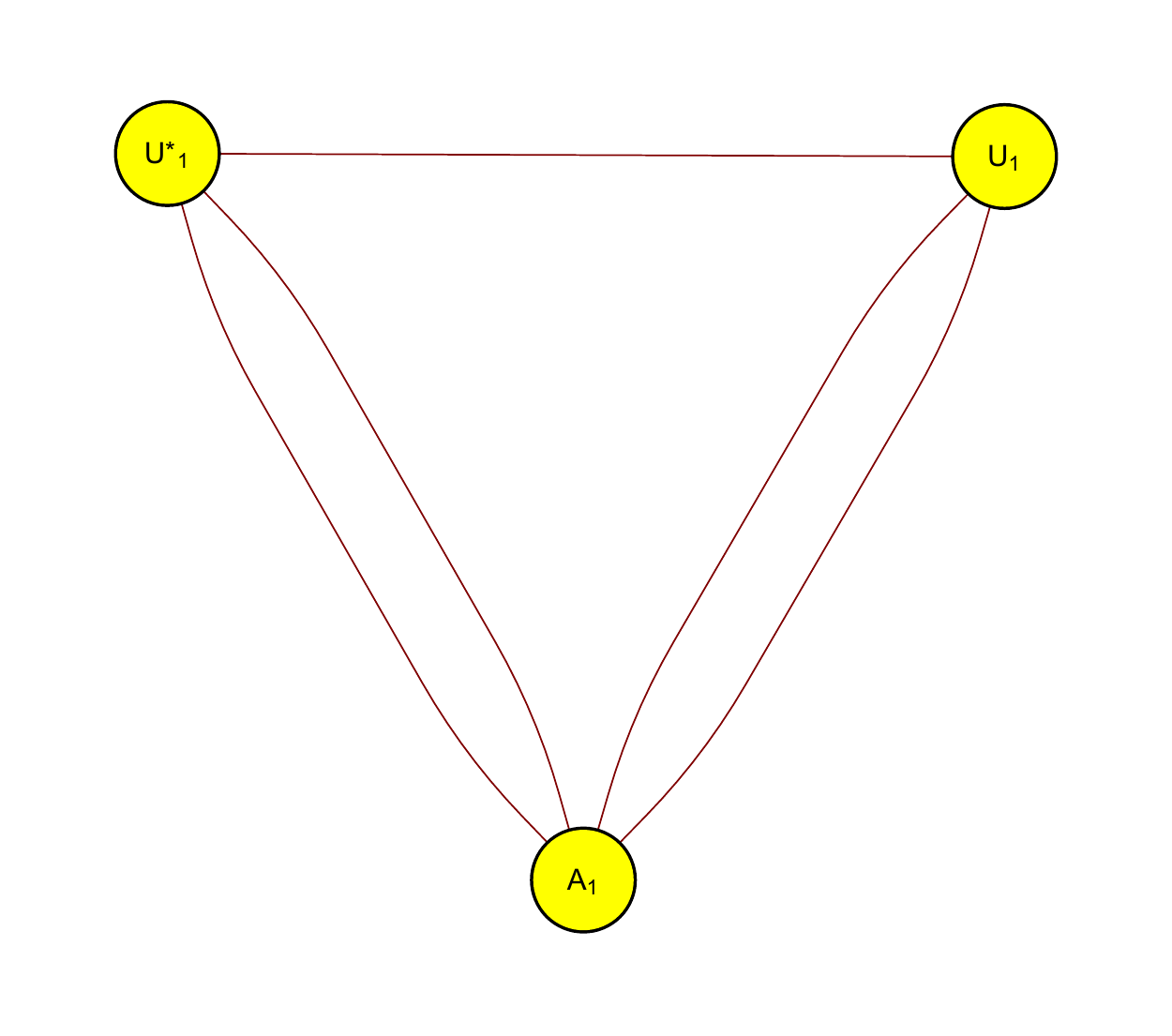} \qquad \includegraphics[align=c, scale=1.0]{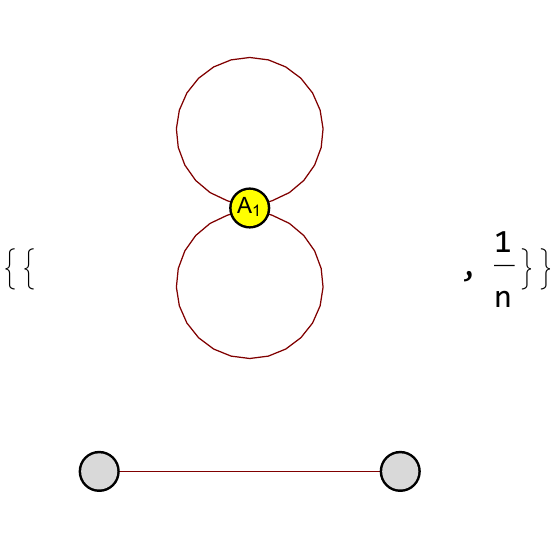}
	
		\caption{Left: Visualization in \texttt{Mathematica} for the tensor network associated to $[\id \otimes \Tr](UAU^*)$. Right: The output tensor network after integrating, with respect to the Haar random unitary matrix $U$, the tensor network on the left. The result consists of two tensor networks, corresponding respectively to $\Tr A$ and to the identity operator on $\mathbb C^n$ (here between two dummy vertices), together with the Weingarten weight $1/n$.}
		\label{fig:m-id-Tr-UU-A-UstarUstar}
	\end{center}
\end{figure}
 The graphical representations of the input tensor network and of the result of the Weingarten integration are presented in Figure \ref{fig:m-id-Tr-UU-A-UstarUstar}. One can easily check that this indeed reproduces  the expression~\eqref{eq:expectationuaustarx}.

\subsection{Bi-partite twirling}
Bi- or multi-partite \emph{twirling} is one of the most useful primitives in quantum information theory, when one has to do with some kind of permutational or rotational symmetry. Introduced by Werner in \cite{werner1989quantum}, it has found many applications in the theory of quantum information. We shall be concerned here with the bi-partite case
\begin{equation}\label{eq:bi-twirling}
\mathcal T(X):= \mathbb E (U \otimes U)X(U^* \otimes U^*),
\end{equation}
where $U \in \mathcal U(d)$ is a Haar-distributed random unitary matrix and $X \in \mathcal M_{d^2}(\mathbb C)$ is a fixed matrix on which the twirling map $\mathcal T$ is acting. The Weingarten computation is more involved in this case, since we are dealing with two copies of the same random unitary matrix (we are in the case $p=2$ of formula \eqref{wgf}).

The analytical computation of the integral in \eqref{eq:bi-twirling} proceeds as follows. First, let $F:\mathbb C^d \otimes \mathbb C^d \to \mathbb C^d \otimes \mathbb C^d$ denote the \emph{flip} (or \emph{swap}) operator, defined by 
$$F x\otimes y = y \otimes x\qquad\textrm{ for }x,y\in\mathbb{C}^d\ .$$
Then, the four choices of the permutations $\alpha,\beta \in \mathcal S_2$ from \eqref{wgf} give four terms, as follows: 

\medskip

\begin{center}
\begin{tabular}{|r||c|c|}
	\hline
$\alpha$ \textbackslash \, $\beta$	& (1)(2) & (12) \\
\hline\hline
(1)(2) & $\Tr(X)I\cdot \frac{1}{d^2-1}$ & $\Tr(XF)I\cdot \frac{-1}{d^3-d}$\\\hline
(12) & $\Tr(X)F\cdot \frac{-1}{d^3-d}$ & $\Tr(XF)F\cdot \frac{1}{d^2-1}$\\\hline
\end{tabular}
\end{center}

\medskip

Putting everything together, we obtain
$$\mathcal T(X) = \frac{1}{d^2-1}\left\{ \left[ \Tr X - \frac{\Tr(XF)}{d} \right]I + \left[ \Tr(XF) - \frac{\Tr X}{d} \right]F \right\}.$$

The \texttt{Mathematica} code used to compute the twirling of an operator is below. 

\begin{lstlisting}
In[1]:= e1 = {{"X", 1, "out", 1}, {"U", 1, "in", 1}};
In[2]:= e2 = {{"U*", 1, "out", 1}, {"X", 1, "in", 1}};
In[3]:= e3 = {{"X", 1, "out", 2}, {"U", 2, "in", 1}};
In[4]:= e4 = {{"U*", 2, "out", 1}, {"X", 1, "in", 2}};
In[5]:= g = {e1, e2, e3, e4}
Out[1]= {{{X, 1, out, 1}, {U, 1, in, 1}}, {{U*, 1, out, 1}, 
        {X, 1, in, 1}}, {{X, 1, out, 2}, {U, 2, in, 1}}, 
        {{U*, 2, out, 1}, {X, 1, in, 2}}}
In[6]:= Eg = integrateHaarUnitary[g, "U", {d}, {d}, d]
Out[2]= {{{{{X,1,out,1},{X,1,in,1}},{{X,1,out,2},{X,1,in,2}},
        {{dummy-U-OUT-1-1,1,in,1},{dummy-U*-IN-1-1,1,out,1}},
        {{dummy-U-OUT-2-1,1,in,1},{dummy-U*-IN-2-1,1,out,1}}},1/(-1+d^2)},
        {{{{X,1,out,1},{X,1,in,1}},{{X,1,out,2},{X,1,in,2}},
        {{dummy-U-OUT-1-1,1,in,1},{dummy-U*-IN-2-1,1,out,1}},
        {{dummy-U-OUT-2-1,1,in,1},{dummy-U*-IN-1-1,1,out,1}}},1/(d-d^3)},
        {{{{X,1,out,1},{X,1,in,2}},{{X,1,out,2},{X,1,in,1}},
        {{dummy-U-OUT-1-1,1,in,1},{dummy-U*-IN-1-1,1,out,1}},
        {{dummy-U-OUT-2-1,1,in,1},{dummy-U*-IN-2-1,1,out,1}}},1/(d-d^3)},
        {{{{X,1,out,1},{X,1,in,2}},{{X,1,out,2},{X,1,in,1}},
        {{dummy-U-OUT-1-1,1,in,1},{dummy-U*-IN-2-1,1,out,1}},
        {{dummy-U-OUT-2-1,1,in,1},{dummy-U*-IN-1-1,1,out,1}}},1/(-1+d^2)}}
\end{lstlisting}

\begin{figure}[htbp]
	\begin{center}
\includegraphics[align=c, scale=0.3]{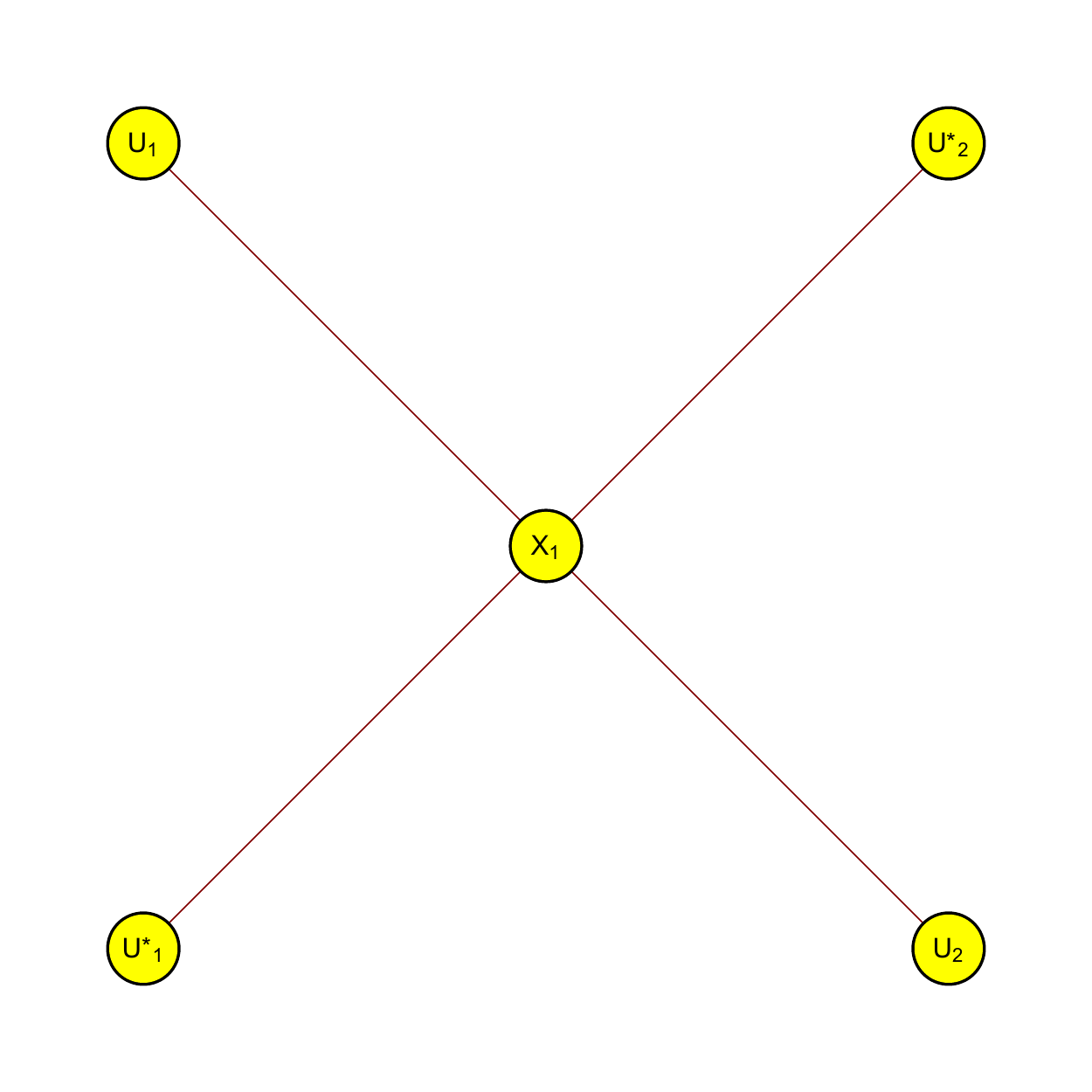} \qquad\qquad\qquad \includegraphics[align=c, scale=0.5]{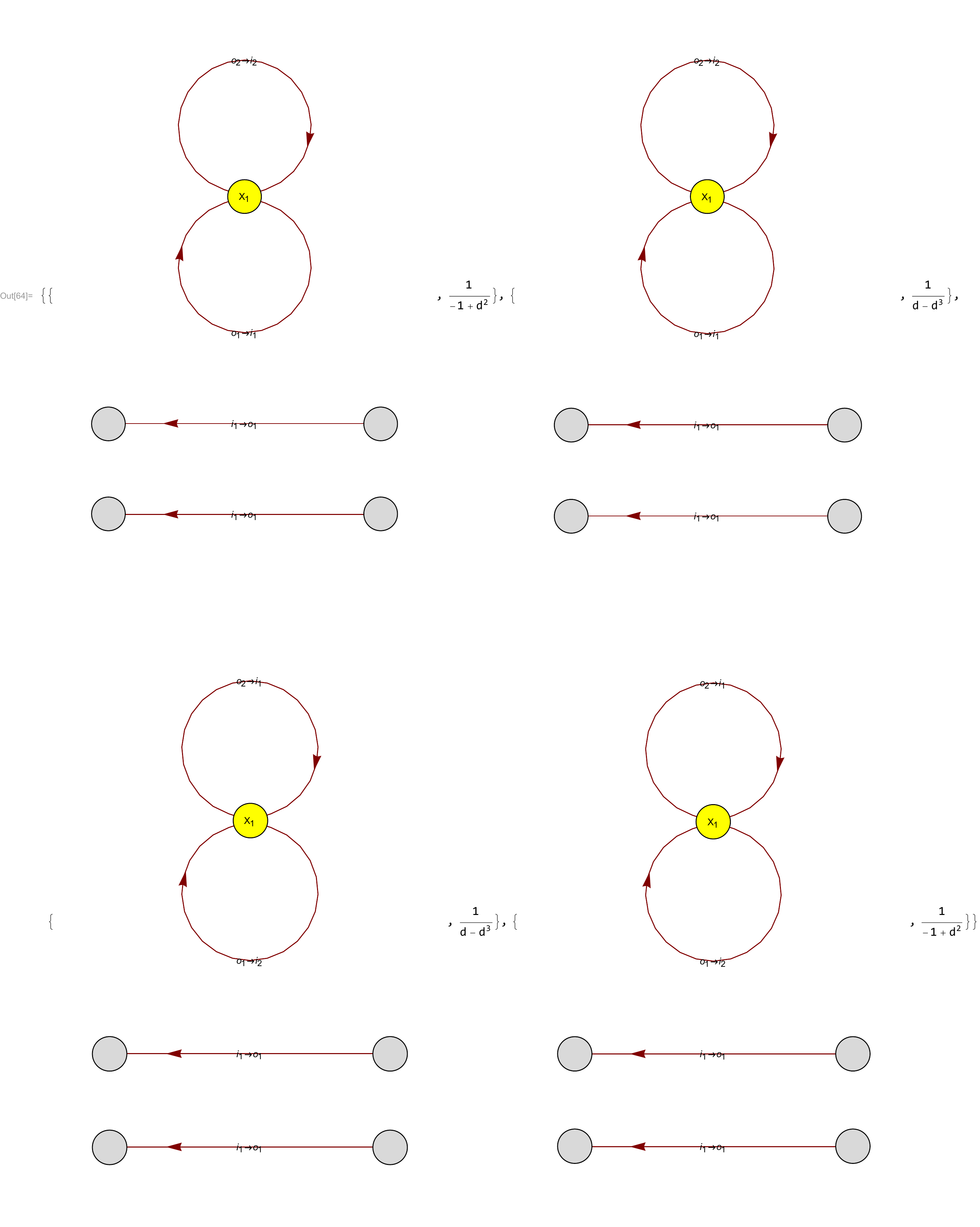}
			\caption{Top: Visualization in \texttt{Mathematica} for the tensor network associated with $(U \otimes U)X(U^* \otimes U^*)$,			generated with \textbf{visualizeTN}[g]. Bottom: The output tensor network after integrating, with respect to the Haar random unitary matrix $U$, the tensor network at the top (figure generated with \textbf{visualizeTN}[Eg,$\{$EdgeLabeling$\rightarrow$ True$\}$]). The result consists of four tensor networks
			and gives the twirl $\mathcal{T}(X)= \mathbb E (U \otimes U)X(U^* \otimes U^*)$ of $X$.	}
		\label{fig:Etwirlx}
	\end{center}
\end{figure}

Note that the output  list of (weighted) tensor networks consists of 4 tensor networks, corresponding, respectively, to the entries $(1,1)$, $(2,1)$, $(1,2)$, $(2,2)$ of the table above.  The input and output are illustrated in Figure~\ref{fig:Etwirlx}.

\subsection{Bell state as input in conjugate random quantum channels}	
\begin{figure}
	\centering
	\includegraphics[width=0.4\linewidth]{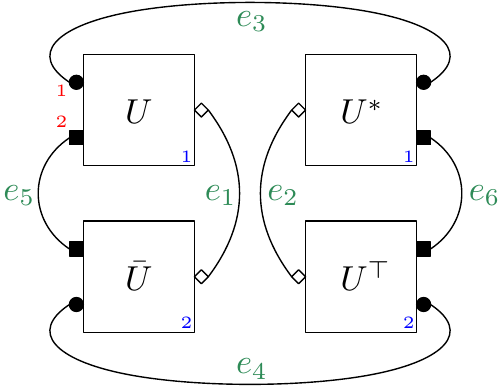}
	\caption{Graphical representation of the scalar $f$ from \eqref{eq:def-f}. Two copies of $U$ and $U^*$ boxed are represented, denoted by \textcolor{blue}{1} and \textcolor{blue}{2} (labels in the bottom-right corner of the boxes). There are two outputs of $U$ (resp.~two inputs of $U^*$), $\mathbb C^n$ (denoted by \textcolor{red}{1}) and $\mathbb C^k$ (denoted by \textcolor{red}{2}), see the labels near the decorations of the first copy fo the $U$ box. The six edges are also identified. }
	\label{fig:overlap}
\end{figure}

In this section, we discuss  an application of \texttt{RTNI} to the theory of random quantum channels. Consider a random quantum channel given by a random isometry $V: \mathbb C^d \to \mathbb C^k \otimes \mathbb C^n$, 
\begin{align*}
\Phi: M_d(\mathbb C) &\to M_k(\mathbb C)\\
X &\mapsto [\mathrm{id}_k \otimes \operatorname{Tr}_n](VXV^*).
\end{align*}
Such random quantum channels were first considered in \cite{hayden2008counterexamples} in relation to the additivity conjecture in quantum information theory, solved in the negative by Hastings \cite{hastings2009superadditivity}. In such considerations, one bounds the minimum output entropy of the tensor product channel $\Phi \otimes \bar \Phi$ by using the overlap
\begin{equation}\label{eq:def-f}
f:=\operatorname{Tr}[\omega_k \cdot [\Phi \otimes \bar \Phi](\omega_d)]
\end{equation}
between the output of the channels acting on a maximally entangled state $\omega_d$ with another maximally entangled state (on the output spaces) $\omega_k$. Here $\bar \Phi$ is the complex conjugate channel. The diagram for the scalar $f$ is given in Figure \ref{fig:overlap}. Recall that the maximally entangled state is defined, in general, by $\omega_n := \Omega_n \Omega_n^*$, with 
\begin{equation}\label{eq:def-maximally-entangled-state}
\mathbb C^d \otimes \mathbb C^d \ni \Omega_n:= \frac{1}{\sqrt n} \sum_{i=1}^n e_i \otimes e_i,
\end{equation}
where $\{e_1, \ldots, e_n\}$ is a given orthonormal basis of $\mathbb C^n$. In \cite{hayden2008counterexamples}, it was shown that, for all (random) isometries $V$, $f \geq d/(kn)$. This bound was improved in \cite{collins2010random}, where it was shown that, in the asymptotic regime where $k$ is fixed, $n \to \infty$, and $d \sim tkn$ for some $t \in (0,1)$,
\begin{equation}\label{eq:limit-overlap}
\lim_{n \to \infty}\mathbb E f = t + \frac{1-t}{k^2}.
\end{equation}

We confirm this computation using our symbolic Weingarten integration routines; note that the corresponding edges are also represented in Figure \ref{fig:overlap}. 

\begin{lstlisting}
In[1]:= e1 = {{"U*", 2, "out", 1}, {"U", 1, "in", 1}};
In[2]:= e2 = {{"U*", 1, "out", 1}, {"U", 2, "in", 1}};
In[3]:= e3 = {{"U", 1, "out", 1}, {"U*", 1, "in", 1}};
In[4]:= e4 = {{"U", 2, "out", 1}, {"U*", 2, "in", 1}};
In[5]:= e5 = {{"U", 1, "out", 2}, {"U*", 2, "in", 2}};
In[6]:= e6 = {{"U", 2, "out", 2}, {"U*", 1, "in", 2}};
In[7]:= g = {e1, e2, e3, e4, e5, e6};
In[8]:= listg = {{g, 1/(d k)}};
In[9]:= Eg = integrateHaarUnitary[listg, "U", {d}, {n, k}, n k]
Out[1]= {{{}, -((d^2 k^2 n)/(d k - d k^3 n^2)) - (d k n^2)/(
d k - d k^3 n^2) + (d k^2 n)/(d k^2 n - d k^4 n^3) + (d^2 k n^2)/(
d k^2 n - d k^4 n^3)}}
In[10]:= overlap = Eg[[1, 2]];
In[11]:= overlap = overlap /. {d -> t n k};
In[12]:= Assuming[t > 0 && k > 1, Limit[overlap, n -> Infinity]]
Out[2]= (1 - t)/k^2 + t
\end{lstlisting}

Note that we obtain first a formula for $\mathbb E f$ at finite $n,d$; we then take the limit $n \to \infty$ to recover \eqref{eq:limit-overlap}. 

%\subsection*{Bell state in conjugate independent channels example}
%Here show that the order of integration does not matter

\subsection{Random tensor networks for holographic duality\label{sec:randomTNSholographic}}

In this section, we illustrate the versatility of the \texttt{RTNI} package on an example taken from \cite{hayden2016holographic}. The authors of \cite{hayden2016holographic} relate the entanglement properties  of a model of random tensors to the AdS/CFT correspondence. In our treatment below, we shall consider a simplified version of their model of random tensors (to be precise, we shall not consider \emph{bulk states} as in~\cite[Section 2]{hayden2016holographic}) that will be analyzed on some example with the help of the \texttt{RTNI} routines. 

\begin{figure}
	\centering
	\includegraphics[align=c,width=0.3\linewidth]{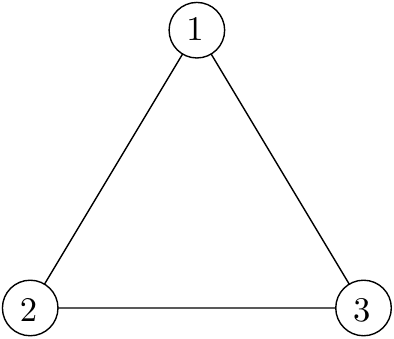} \qquad\qquad\qquad \includegraphics[align=c,width=0.3\linewidth]{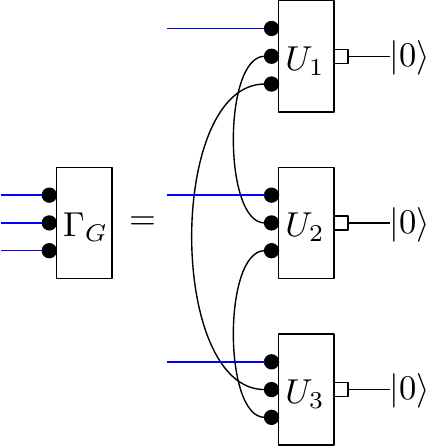}
	\caption{A random tensor (right) associated to a triangle (left).}
	\label{fig:tn-triangle}
\end{figure}

To a given unoriented simple graph $G = (V,E)$ on $n=|V|$ vertices, we shall associate an $n$-tensor $\Psi_G \in (\mathbb C^d)^{\otimes n}$, as follows. For a vertex $x \in V$ of degree $\mathrm{deg}_x$, consider a random pure state (unit vector)
$$(\mathbb C^d)^{\otimes (1+\mathrm{deg}_x)} \ni |V_x \rangle  := U_x |0\rangle,$$
where $U_x$ is a Haar-distributed random unitary matrix and $|0\rangle$ is a fixed vector; moreover, assume that the family $\{U_x\}_{x \in V}$ is independent. Among the $\mathrm{deg}_x+1$ tensor legs of $|V_x\rangle$, the first one is a \emph{dangling edge}, while the $\mathrm{deg}_x$ remaining ones will be contracted with the neighboring vertices to give the state $\Gamma_G$ (see Figure \ref{fig:tn-triangle} for the case of a triangle graph):
$$\Psi_G  := \left[ \bigotimes_{e \in E} \langle \Omega_e| \right] \left[ \bigotimes_{x \in V} |V_x\rangle \right],$$
where $| \Omega \rangle \in \mathbb C^d \otimes \mathbb C^d$ is the maximally entangled state from \eqref{eq:def-maximally-entangled-state}.
%\begin{equation}\label{eq:def-maximally-entangled}
% \ni |\Omega \rangle := \frac{1}{\sqrt d} \sum_{i=1}^d e_i \otimes e_i.
%\end{equation}

For a fixed (possibly empty) subset $A \subseteq V$, consider the marginal
$$\rho_{G;A} := [\mathrm{id}_{A} \otimes \Tr_{\bar A}] \left(\Psi_G \Psi_G^*\right) \in \mathcal M_{d^{|A|}}(\mathbb C).$$
It is important to note that neither $\Psi_G$ nor $\rho_{G;A}$ are normalized quantum states; this is due to the fact that the tensor contractions implemented by the maximally entangled states $|\Omega_e\rangle $ are not isometries. We define the normalized version (we drop the index $G$ to make the notation lighter)
$$ \hat \rho_A := \frac{\rho_{A}}{\Tr \rho_{A}},$$
and we compute, as in \cite{hayden2016holographic}, its $2$-Renyi entropy
$$\exp(-S_2(\hat \rho_{A})) = \Tr (\hat \rho_{A}^2) = \frac{\Tr \rho_{A}^2}{(\Tr \rho_{A})^2} = \frac{\Tr[(\rho_A \otimes \rho_A) F]}{\Tr[\rho_A \otimes \rho_A]} = \frac{\Tr[(\rho \otimes \rho) F_{AA'}]}{\Tr[\rho \otimes \rho]},$$
where $F (x \otimes y) = y \otimes x$ is the \emph{flip operator}, $\rho = \Psi \Psi^* = \rho_\emptyset$ is the (global) density operator,
and $F_{AA'}=F\otimes \mathrm{id}_{V\backslash A}\otimes\mathrm{id}_{V'\backslash A'}$ denotes the flip operator acting on the sites associated with~$A$ (respectively its copy $A'$). In \cite{hayden2016holographic} it is argued that both the numerator and the denominator of the expression above concentrate around their average value (with respect to the randomness in the Haar unitaries $\{U_x\}_{x \in V}$), denoted in general by
\begin{align}
\bar Z_A := \mathbb E \Tr[(\rho \otimes \rho)\mathcal F_A]\ .\label{eq:zaaverageexpr}
\end{align}
In~\cite[Section 2.2]{hayden2016holographic}, the value  of averages of the form~\eqref{eq:zaaverageexpr} was shown to be expressible -- to lowest order in the limit~$d\rightarrow\infty$ -- in terms of the minimum energy associated with an Ising model with classical spins on the vertices of the graph~$G$, and a magnetic field determined by the sites~$A$ of interest. More precisely, the following expression was derived:
\begin{align}
\lim_{d \to \infty} \frac{-\log \bar Z_A}{\log d} &= \min_{s \in \{\pm 1\}^V} \left\{  \sum_{x \in V} 2(\mathrm{deg}_x+1) - \frac 1 2 \left[ \sum_{e = (x,y) \in E} (s_x s_y-1) + \sum_{x \in V} (h_xs_x+3) \right]\right\}\nonumber\\
&= \frac{|V|}{2} + \frac{9|E|}{2} - \frac 1 2\max_{s \in \{\pm 1\}^V}  \left[ \sum_{e = (x,y) \in E} s_x s_y + \sum_{x \in V} h_xs_x \right]\ ,\label{eq:dinfinityzalogde}
\end{align}
where 
$$h_x = \begin{cases}
-1,&\qquad \text{ if } x \in A\\
+1,&\qquad \text{ if } x \notin A.
\end{cases}$$
Obviously, in the case where $A = \emptyset$ (which corresponds to the normalizing denominator for the state $\rho_A$), the unique optimum is $s \equiv 1$, and we obtain
\begin{align}
\lim_{d \to \infty} \frac{-\log \bar Z_\emptyset}{\log d} = 4|E|\ .\label{eq:nomarginalexamplex}
\end{align}

\begin{figure}
	\centering
	\includegraphics[align=c,scale=0.3]{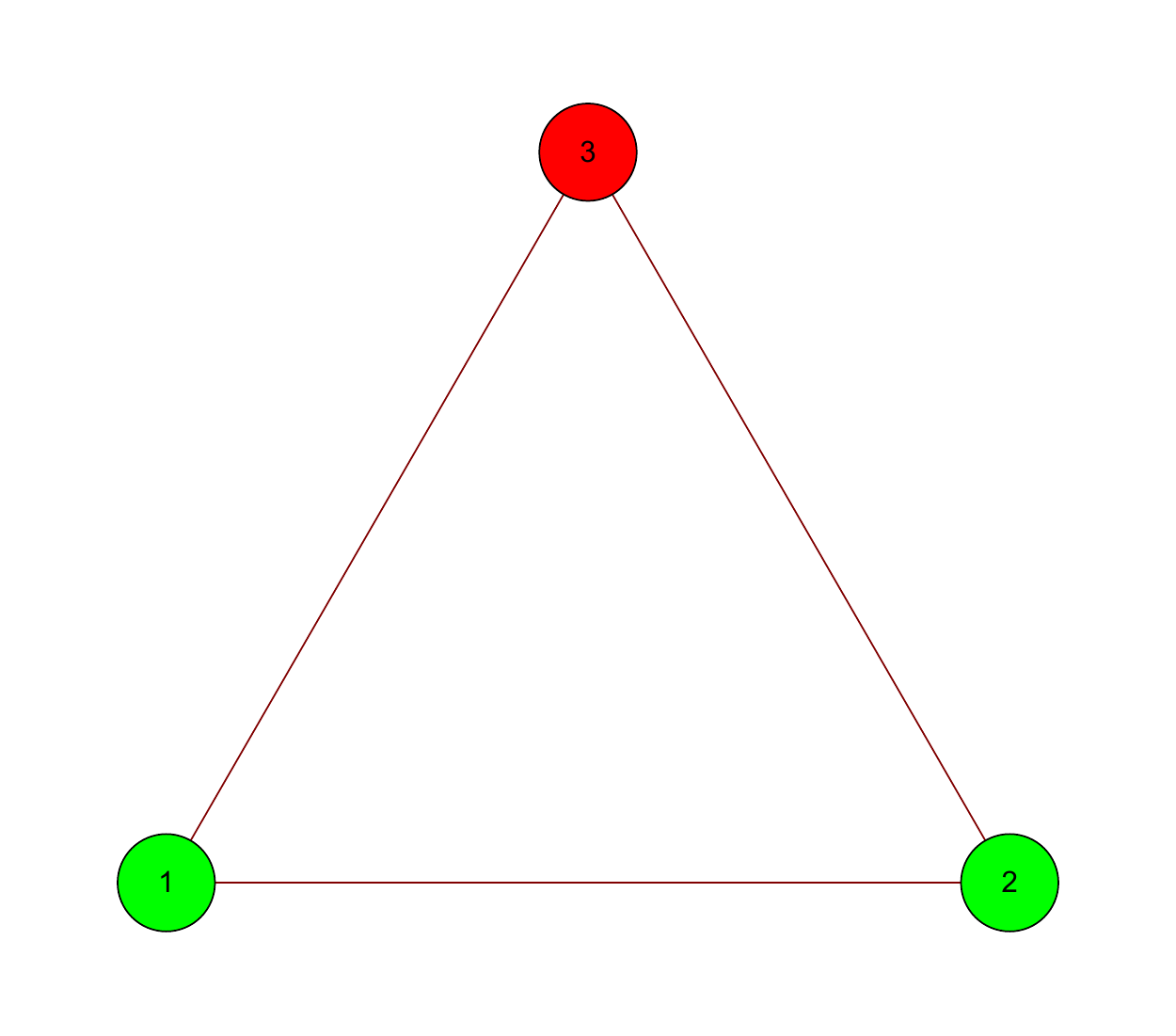} \quad
	\includegraphics[align=c,scale=0.5]{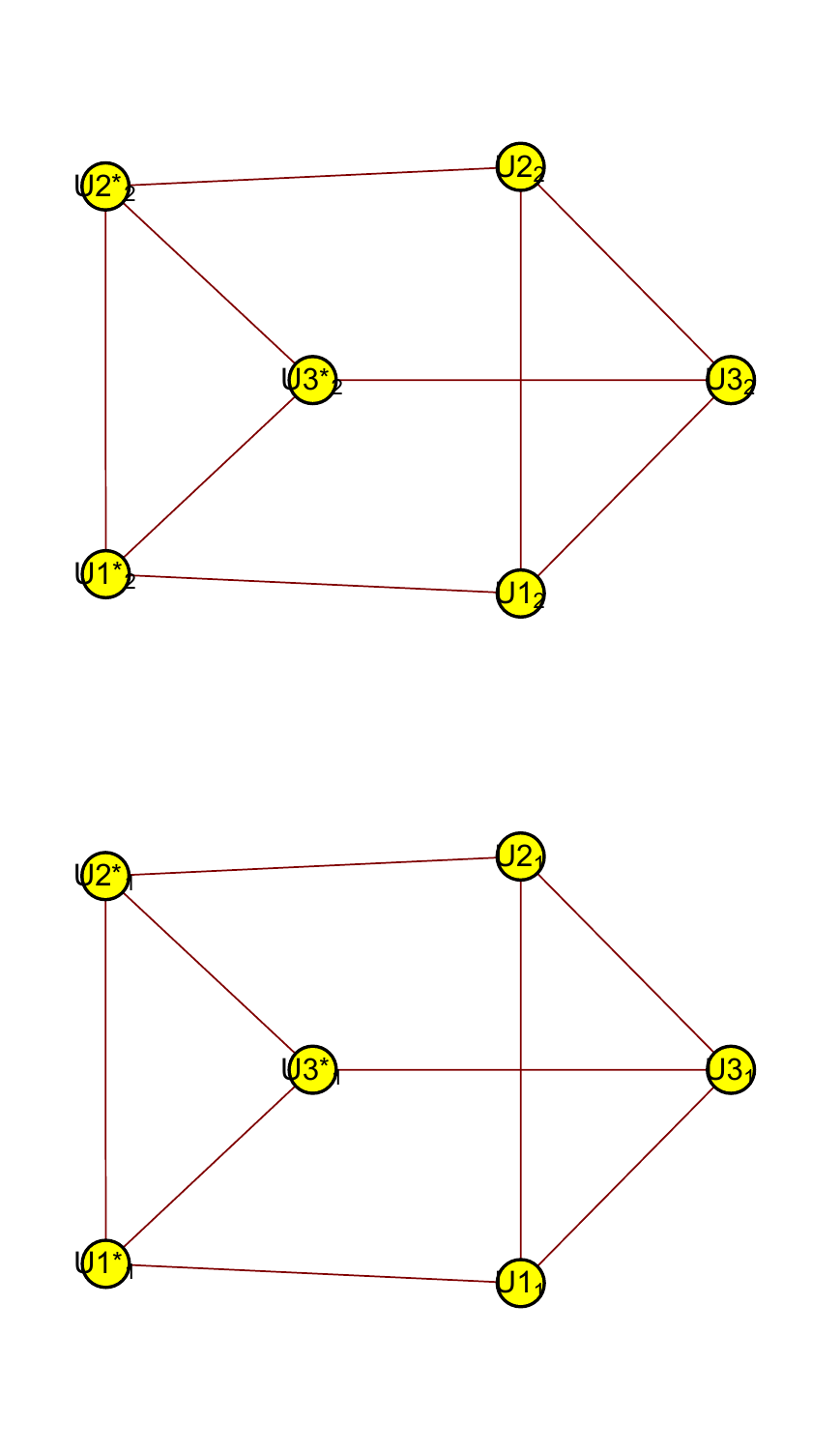} \quad
	\includegraphics[align=c,scale=0.5]{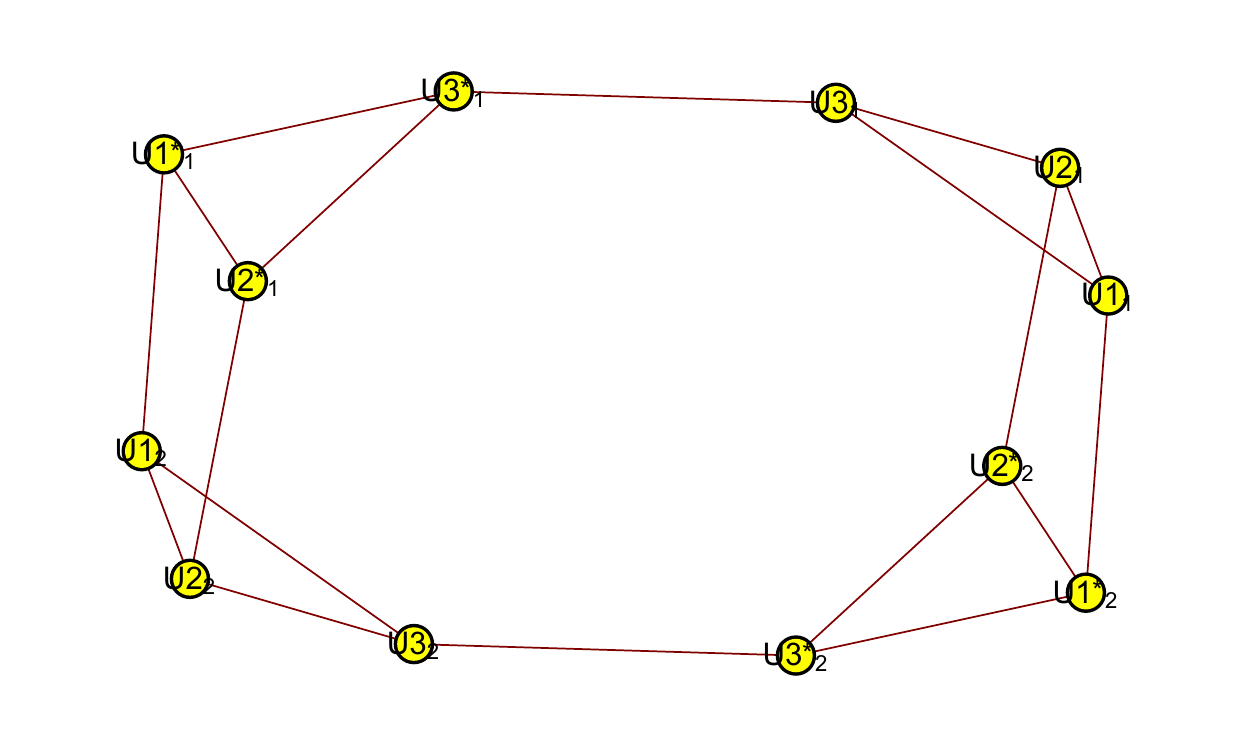}
	\caption{From left to right: the $\{1,2\}$-marginal of the triangle graph; the tensor network corresponding to $\Tr[\rho \otimes \rho]$; the tensor network corresponding to $\Tr[\rho_{\{1,2\}}^2]$.}
	\label{fig:RTNI-triangle}
\end{figure}

Here we use the \texttt{RTNI} package to verify the validity of~\eqref{eq:dinfinityzalogde} and~\eqref{eq:nomarginalexamplex} for simple example of graphs~$G$~\footnote{In~\cite{hayden2016holographic}, expressions~\eqref{eq:dinfinityzalogde} and~\eqref{eq:nomarginalexamplex} are subsequently used to derive Ryu-Takayanagi-type formulas for the R\'enyi entropy. We do not consider this here as this derivation appears to require additional assumptions on connectivity properties of the graph~$G$, guaranteeing the minimum-energy considerations have a single domain wall (unlike in triangle graph example considered here). }. In the case of the triangle graph~$G_{\textrm{triangle}}$ shown in Figure~\ref{fig:tn-triangle}, with a single point marginal $A=\{1,2\}$, we obtain again that the optimizing configuration is $s_x=1$ for all $x\in V$ and thus we obtain from~\eqref{eq:nomarginalexamplex} and~\eqref{eq:dinfinityzalogde}
\begin{align}
\lim_{d \to \infty} \frac{-\log \bar Z_{\emptyset}}{\log d} &= 12\\
\lim_{d \to \infty} \frac{-\log \bar Z_{\{1,2\}}}{\log d} &= 13\ .
\end{align}
This result is confirmed by the following \texttt{RTNI} exact, non-asymptotic, computation (see also Figure \ref{fig:RTNI-triangle}). Note that the pre-factor $d^{-2|E|}$ appears below to take into account the normalization of the maximally entangled states \eqref{eq:def-maximally-entangled-state}, which is not encoded directly into the tensor network.
\begin{lstlisting}
In[1]:= g = Graph[{1 <-> 2, 2 <-> 3, 3 <-> 1}];
In[2]:= edgeNormalization = d^(-2 EdgeCount[g]);
In[3]:= marginal = {1, 2};
In[4]:= tn = buildTNfromGraph[g, marginal];
In[5]:= Z0 = edgeNormalization integrateAllUs[tn[[1]], d, d][[1, 2]] // 
  FullSimplify
Out[1]= (3 + (-2 + d) d)/(d^7 (1 + d) (1 + (-1 + d) d)^3)
In[6]:= Limit[-Log[Z0]/Log[d], d -> Infinity]
Out[2]= 12
In[7]:= ZA = edgeNormalization integrateAllUs[tn[[2]], d, d][[1, 2]] // 
  FullSimplify
Out[3]= (1 + d^2)/(d^8 (1 + d) (1 + (-1 + d) d)^3)
In[8]:= Limit[-Log[ZA]/Log[d], d -> Infinity]
Out[4]= 13
\end{lstlisting}

Here the routine~\textbf{buildTNfromGraph} creates two tensor networks starting from a graph~$G$ and a subset~$A\subset V$ of vertices. The tensor networks represent~$\Tr[\rho\otimes\rho]$ and $\Tr[\rho_A^2]$, respectively (see Fig.~\ref{fig:RTNI-triangle}). The corresponding code and additional examples for e.g., 2D~grid graphs are provided in a separate \texttt{Mathematica} notebook.

\subsection{Example computation of expectation values of matrix products\label{sec:multinomialexpectationvalueexamples}}
Here we illustrate the use of \textbf{MultinomialexpectationvalueHaar}. 
We  use the routine to obtain
\begin{align}
\mathbb E \Tr[XUYU^*] &=\Tr[X]\Tr[Y]/d\\
\mathbb E XUYU^T&=0\\
\mathbb E XU^*YU^T&=XY^T/d\\
\mathbb E \Tr[XUY\overline{U}]&=\Tr[XY^T]/d\ 
\end{align}
 for two $d\times d$-matrices~$\alpha,\beta$ and a Haar-random unitary~$U$  of size~$d$.
\begin{lstlisting}
In[1]:= MultinomialexpectationvalueHaar[d,{1,2},{X,Y},True]
Out[2]= (Tr[X] Tr[Y])/d
In[3]:= MultinomialexpectationvalueHaar[d,{1,3},{X,Y},False]
Out[4]= 0
In[5]:= MultinomialexpectationvalueHaar[d,{2,3},{X,Y},False]
Out[6]= X.Transpose[Y]/d
In[7]:= MultinomialexpectationvalueHaar[d,{1,4},{X,Y},True]
Out[8]= Tr[X.Transpose[Y]]/d
\end{lstlisting}
For two additional matrices $V,W$, we can compute
for example
\begin{align}
\mathbb E VUWU^*XU^TY\overline{U}&=
-\frac{1}{d^4-5d^2+4} VY^TXW^T-\frac{\Tr(W)}{d^4-5d^2+4}VY^TW\\
&\quad
-\frac{\Tr(Y)}{d^4-5d^2+4}VXW^T-\frac{\Tr(W)\Tr(Y)}{d^4-5d^2+4}VX
\end{align}
using the command
\begin{lstlisting}
In[8]:= MultinomialexpectationvalueHaar[d,{1,2,3,4},{V,W,X,Y},False]
Out[9]= -(V.Transpose[Y].X.Transpose[W]/(4-5 d^2+d^4))
        -(V.Transpose[Y].X Tr[W])/(4-5 d^2+d^4)
        -(V.X.Transpose[W] Tr[Y])/(4-5 d^2+d^4)
        -(V.X Tr[W] Tr[Y])/(4-5 d^2+d^4)
\end{lstlisting}

\bigskip

\noindent\textit{Acknowledgments.} I.N.'s research has been supported by the ANR projects {StoQ} {ANR-14-CE25-0003-01} and {NEXT} {ANR-10-LABX-0037-NEXT}. I.N.~and M.F.~acknowledge the hospitality of the TU M\"unchen, where part of this work was conducted,
and are both supported by the PHC Sakura program (project number: 38615VA). R.K.~acknowledges support by DFG project no.~K05430/1-1 and  the Technical University of Munich -- Institute for Advanced Study, funded by the German Excellence Initiative
and the European Union Seventh Framework Programme under grant agreement no.~291763.
 M.F. was financially supported by JSPS KAKENHI Grant Number JP16K00005.
 
\bibliography{ref}{}
\bibliographystyle{alpha}

\appendix

\section{User interface of \texttt{Python} package}
In this section, we show how the \texttt{Python} package works, by using the example in Section \ref{sec:several_tensor}: $[\id \otimes \Tr](UAU^*)$. First, the following code creates the initial graph and visualizes it.
\begin{lstlisting}
In[1]: from IHU_source import *
In[2]: e1 = [["A", 1, "out", 1], ["U", 1, "in", 1]]
In[3]: e2 = [["A", 1, "out", 2], ["U", 1, "in", 2]]
In[4]: e3 = [["U*", 1, "out", 1], ["A", 1, "in", 1]]
In[5]: e4 = [["U*", 1, "out", 2], ["A", 1, "in", 2]]
In[6]: e5 = [["U", 1, "out", 2], ["U*", 1, "in", 2]]
In[7]: g = [e1, e2, e3, e4, e5]
In[8]: gw = [g,1]
In[9]: visualizeTN(gw)
\end{lstlisting}

%\begin{lstlisting}
%In[1]: from IHU_source import *
%In[2]: e1 = [["A", 0, "L", 0], ["U", 0, "R", 0]]
%In[3]: e2 = [["A", 0, "L", 1], ["U", 0, "R", 1]]
%In[4]: e3 = [["U*", 0, "L", 0], ["A", 0, "R", 0]]
%In[5]: e4 = [["U*", 0, "L", 1], ["A", 0, "R", 1]]
%In[6]: e5 = [["U", 0, "L", 1], ["U*", 0, "R", 1]]
%In[7]: g = [e1, e2, e3, e4, e5]
%In[8]: gw = [g,1]
%In[9]: visualizeGraph(gw)
%\end{lstlisting}
Note that the input line 1 shows that source codes are imported from the file ``IHU\_source.py'',  which imports source codes from ``WFG\_source''.  

\begin{figure}
	\centering
	\includegraphics[width=0.5\linewidth]{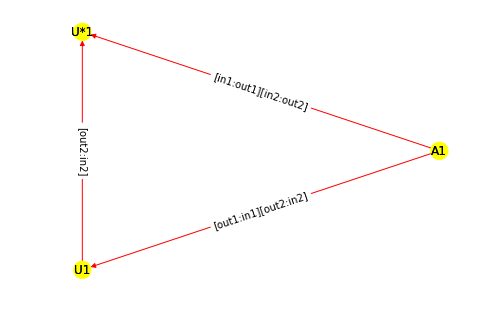}
	\caption{The visualization of the input tensor network, which  is followed by the initial weight $1$}
	\label{fig:p_before}
\end{figure}

One difference (compared to the \texttt{Mathematica} code) is in the input line 8, where the weight is explicitly specified, here as~$1$. If one wants to use symbolic number for the initial weight, it must be defined as described  below using the \textbf{symbols} command.
% In addition, there is the following difference in syntax between the two packages:
%\vspace{1ex}
%\begin{itemize} 
%\item \Mathematica: ``out'' respectively  ``in'' 
%\item \Python:   ``L'' respectively ``R'' 
%\end{itemize}
%\vspace{1ex}
%In other words, the ``left side'' of matrices corresponds to the output space and the ``right side'' to the input. 
 Figure~\ref{fig:p_before} shows the visualization of the initial tensor network before taking average. 
In this figure, matrices are represented as yellow disks and red arrows carry information about how matrices are connected. For example, consider the arrow from ``U*1'' to ``A1''. The information ``[in1:out1][in2:out2]'' shows that this is a double edge, where the first bracket come from the edge ``e3'' and the second the edge ``e4''.

\begin{figure}
	\centering
	\includegraphics[width=0.5\linewidth]{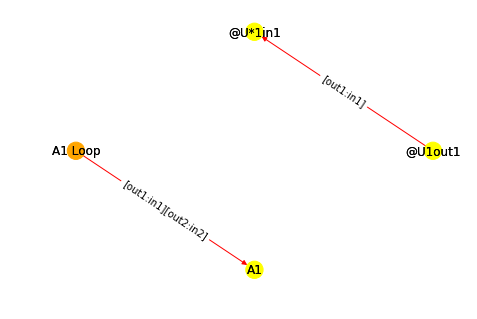}
	\caption{The visualization of the output tensor network, which is followed by the new weight $1/n$}
	\label{fig:p_after}
\end{figure}

Next, the following code averages the input graph over the unitary group and visualize it.
%In[1]: from IHU_source import *
%In[2]: k,n = symbols('k n')
%In[3]: e1 = [["A", 0, "L", 0], ["U", 0, "R", 0]]
%In[4]: e2 = [["A", 0, "L", 1], ["U", 0, "R", 1]]
%In[5]: e3 = [["U*", 0, "L", 0], ["A", 0, "R", 0]]
%In[6]: e4 = [["U*", 0, "L", 1], ["A", 0, "R", 1]]
%In[7]: e5 = [["U", 0, "L", 1], ["U*", 0, "R", 1]]
%In[8]: g = [e1, e2, e3, e4, e5]
%In[9]: gw = [g,1]
\begin{lstlisting}
In[10]: k,n = symbols('k n')
In[11]: rm = ["U",[n,k],[n,k],n*k]
In[12]: Eg = integrateHaarUnitary(gw,rm)
In[13]: print(Eg)
In[14]: visualizeTN(Eg)
Out[1]: [[[[['@U*', 1, 'in', 1], ['@U', 1, 'out', 1]], 
[['A', 1, 'out', 1], ['A', 1, 'in', 1]], 
[['A', 1, 'out', 2], ['A', 1, 'in', 2]]], 1/n]]
\end{lstlisting}
Note that in the input line 2, $k$ and $n$ are declared to be symbols explicitly. One can define other symbols in the same way. 
The algorithm calculates the average in input line 11 and outputs the result in the next line as shown in output line 1. 
The output is again a list, where the outer bracket are there in case the output is a sum of several tensor networks with corresponding weights.
In this example, since we only have a pair of $U$ and $U^*$, the average can be written as a single pair of a tensor network and a weight; the weight is now $1/n$ and the remainder of the output specifies the tensor network. The mark ``@'' corresponds to ``dummy'' in the \texttt{Mathematica} package. 
A visualization of the averaged tensor network is given in Figure~\ref{fig:p_after}. 
Loops are represented by orange disks. In this example, one orange disk is connected to the yellow disk labeled as ``A1'' by the red arrow with the information ``[out1:in1][out2:in2]''. This means $\Tr  A$.

\section{Generating Weingarten functions}\label{sec:GW}
In this section, we provide a brief explanation of  how the \texttt{Python} algorithm for generating Weingarten functions works,  and show how to use it.

\subsection{How the algorithm works}\label{sec:MN-rule}
The formula for  Weingarten functions \eqref{eq:wg} consists of two ingredients: characters of symmetric groups and Schur polynomials (i.e., characters of the unitary group) evaluated at the identity. Computation of the latter is straightforward based on the formula~\eqref{eq:schur}. The calculation of the former is rather complicated. We use the Murnaghan--Nakayama rule~ \cite{Murnaghan, Nakayama1, Nakayama2}. There are recursive and non-recursive versions, and we choose to use a  recursive one because it uses only Young diagrams, while the non-recursive ones requires manipulation of Young Tableaux. We refer to~\cite{Bernstein} for a more detailed analysis of the problem of computing character tables of symmetric groups.

To make our paper self-contained, let us explain this recursive formula.  The character table of the symmetric group $\cS_p$ is a square table, with entries $\chi(\alpha,\beta)$ indexed by two ordered partitions~$\alpha,\beta$ of $p$. Here $\alpha$ denotes a  Young diagram specifying an irreducible representation, and $\chi(\alpha,\beta)=\chi_\alpha(\pi)$ is the character evaluated for any permutation~$\pi\in\cS_p$ with cycle type~$\beta$. A border strip is a connected set of boxes contained in a Young diagram. Now, fix $\alpha = (\alpha_1, \ldots, \alpha_m)$ and $\beta = (\beta_1,\ldots,\beta_n)$ and take the Young diagram of the form $\alpha$ and find all border strips within $\alpha$, denoted by $\gamma = (\gamma_1, \ldots, \gamma_m)$, such that 
\begin{enumerate}
\item $\displaystyle \sum_{i=1}^m \gamma_i = \beta_1$
\item removing $\gamma$ from the Young diagram yields another Young diagram
\item $\gamma$ does not contain $2\times 2$ square. 
\end{enumerate}
We call any such border strip {\em valid}. From the second condition above it follows that the boxes of a valid border strip occupy the right-hand-side of each row of the Young diagram. 

\begin{figure}
	\centering
	\includegraphics[width=0.4\linewidth]{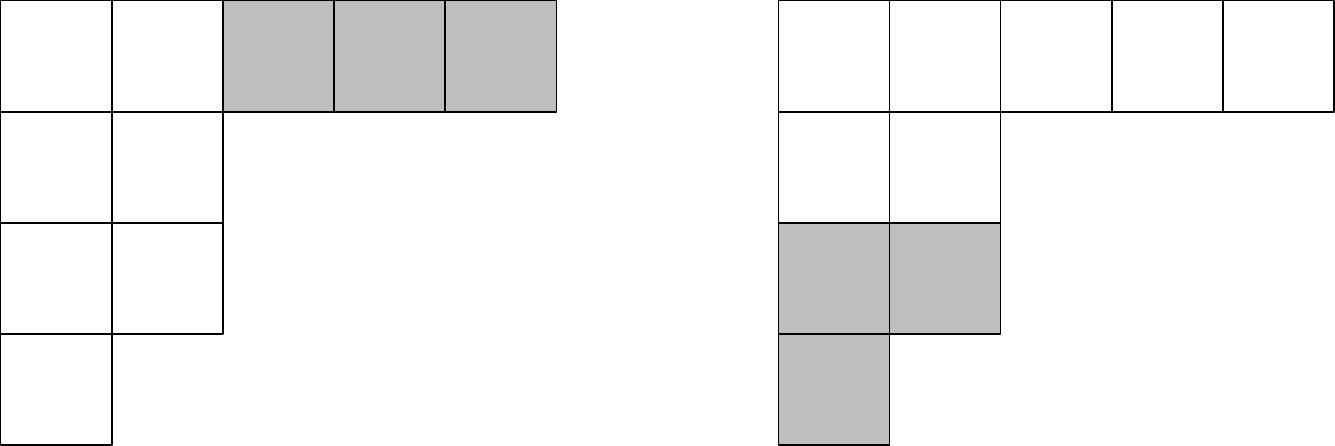}
	\caption{Two valid border strips are shaded. The associated heights as defined in Eq.~\eqref{eq:heightdefinition} are 0 and 1, respectively. We will denote the border strips as $(3,0,0,0)$ and $(0,0,2,1)$, respectively. These are the only valid border strips of size 3 contained in the Young diagram of shape $(5,2,2,1)$.}
	\label{fig:bs_good}
\end{figure}

\begin{figure}
	\centering
	\includegraphics[width=0.4\linewidth]{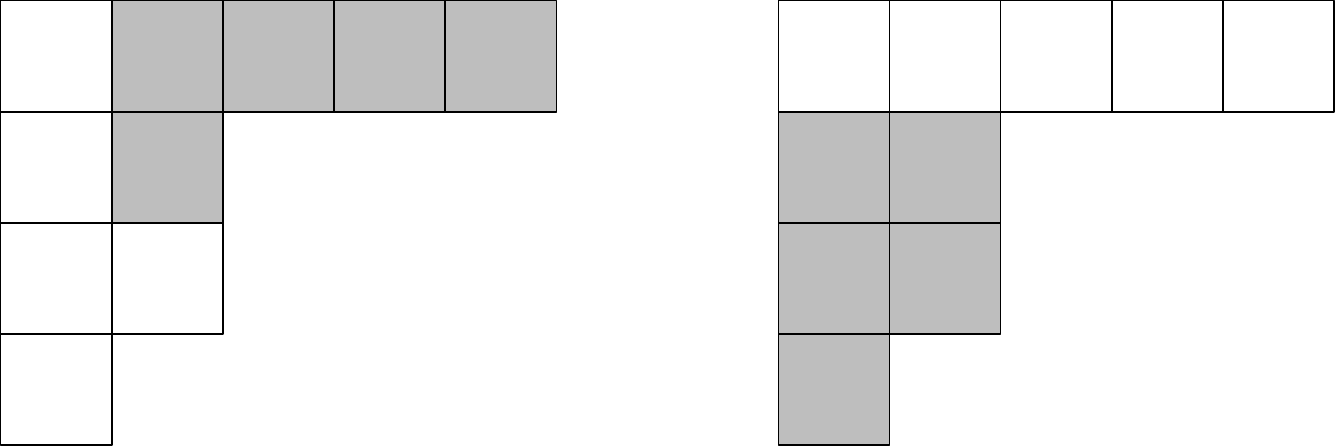}
	\caption{The shaded boxes do not constitute valid border strips: the one on the left
	does not result in a Young diagram when it is removed. The one on the right contains a $2\times 2$~square.	}
	\label{fig:bs_bad}
\end{figure}

Let us consider some examples. Suppose $\alpha = (5,2,2,1)$. One can find two valid border strips of size $3$ (Figure \ref{fig:bs_good}) but none of size $5$ (Figure \ref{fig:bs_bad}).

Then, calculate the height~$h(\gamma)$  of each border strip~$\gamma$, which is defined as 
\begin{align}
h(\gamma) = (\text{\# of rows in }\gamma) -1 .\label{eq:heightdefinition}
\end{align}
(In the notation introduced in Figure~\ref{fig:bs_good}, the number of rows is the number of non-zero entries of~$\gamma$.) The quantity~$\chi(\alpha,\beta)$ then is given by the recursive formula
\begin{align}
\chi(\alpha,\beta) = \sum_{\gamma} (-1)^{h(\gamma)} \, \chi(\alpha - \gamma, (\beta_2,\ldots, \beta_n))\ \label{eq:mgrulerec}
\end{align}
where the sum is over all valid border strips~$\gamma$ for the diagram~$\alpha$. 
Here the expression~$\alpha - \gamma$ refers to the Young diagram obtained by removing the boxes of~$\gamma$ from~$\alpha$. 

Let us illustrate this using the example of the diagram~$\alpha=(5,2,2,1)$ considered above. One can compute for example
\begin{align}
\chi ((5,2,2,1),(3,2,2,2,1)) = (-1)^0 \, \chi ((2,2,2,1),(2,2,2,1)) + (-1)^1 \, \chi((5,2),(2,2,2,1))
\end{align}
where the sume was taken over the valid border strips $(3,0,0,0)$ and $(0,0,2,1)$. By contrast,
\begin{align}
\chi ((5,2,2,1),(5,4,1)) = 0
\end{align}
because there is no border strip of size $5$ which can be removed from the Young diagram of~$(5,2,2,1)$.

Successive application of the recursive rule~\eqref{eq:mgrulerec} 
leads to a linear combination of $\chi((1),(1))$ and $\chi((),())$, which are both defined to be $1$.
However, our program uses the rule~\eqref{eq:mgrulerec} just once, and computes the character table of $\cS_p$ based on all the character tables of $\cS_1,\ldots,  \cS_{p-1}$.
This is why if one computes Weingarten functions for $\cS_p$, the program  automatically generates the character tables of $\cS_1, \ldots,\cS_p$.

\subsection{How to use the \texttt{Python} \texttt{RTNI} package for generating Weingarten functions}
One can generate Weingarten functions using the \texttt{Python} package. Example code is given here:
\begin{lstlisting}
In[1]: from WFG_source import *
In[2]: k = 3
In[3]: display='yes'
In[4]: record = 'yes'
In[5]: weigartenFunctionGenerator(k,display,record)
Out[1]: [[[1, 1, 1], (n**2 - 2)/(n*(n**4 - 5*n**2 + 4))], 
[[2, 1], -1/(n**4 - 5*n**2 + 4)], [[3], 2/(n*(n**4 - 5*n**2 + 4))]]
\end{lstlisting}
where $k$ is the size of the symmetric group. The output gives the Weingarten function for dimension~$n$ (i.e., for the unitary group on~$\mathbb{C}^n$)  as a list of pairs $(\alpha,\Wg_n(\pi_\alpha))$ where $\alpha $ (an ordered partition of $k$) is the cycle type and $\pi_\alpha\in \cS_k$ is any permutation with cycle type~$\alpha$. 

By choosing 'yes' or 'no', one can specify if the algorithm displays the result and if it should be recorded as "functions$k$" in the folder "Weingarten". In case the folder is missing, it is created automatically. In the example, $k=3$ and Weingarten functions for $\cS_3$ and $\mathcal U(n)$ are calculated in the output, being represented as a list. 
For example, the Weingarten function for the partition (cycle type)~$[1,1,1]$ is $(n^2 - 2)/(n(n^4 - 5n^2 + 4))$.

Moreover, one can generate only the character tables of symmetric groups with the following code.
\begin{lstlisting}
In[1]: from WFG_source import *
In[2]: k = 3
In[3]: display='yes'
In[4]: characterTableGenerator(k,display)
Out[1]: {'[1, 1, 1][1, 1, 1]': 1, '[1, 1, 1][2, 1]': -1, '[1, 1, 1][3]': 1, 
'[2, 1][1, 1, 1]': 2, '[2, 1][2, 1]': 0, '[2, 1][3]': -1, '[3][1, 1, 1]': 1, 
'[3][2, 1]': 1, '[3][3]': 1}
\end{lstlisting}
Here, $k$ is the size of the symmetric group, as before, and it is possible to choose if the algorithm shows the result or not. The result will be recorded automatically in the folder ``SGC'', which is created in case it is missing. 
In this example, $k=3$ and each of the output tuples stands for an element of character table. For example, by using the notation $\chi$ in Section \ref{sec:MN-rule}, $\chi ((2,1),(3)) =-1$. All the above data is stored as ``.pkl'' files, which can be translated to text files by the program called ``pkl2text''.

\end{document}